\documentclass[subscriptcorrection,upint,varvw,barcolor=Goldenrod3,mathalfa=cal=euler,balance,hyphenate,french,pdf-a,nolists,table,xcdraw]{asmejour} %

\usepackage[switch]{lineno} 

\usepackage[noend]{algpseudocode}
\usepackage{algorithm}
\usepackage{algorithmicx}
\usepackage{etoolbox}

\usepackage{enumitem}

\hypersetup{%
	pdfauthor={Saeid Bayat},                       		   	
	pdftitle={Multi-split configuration design for fluid-based thermal management systems},                  	
	pdfkeywords={Thermal Management System Design, Design Synthesis, Optimization, Optimal Flow Control, Knowledge Extraction, Graph Modeling},
	pdfsubject = {Describes the asmejour LaTeX template},			
	pdflicenseurl={https://ctan.org/pkg/asmejour},
}

\AtBeginEnvironment{algorithmic}{\lineskip0pt}

\usepackage{enumitem}
\usepackage{graphicx}
\setlist[enumerate,1]{label=\arabic*}
\setlist[enumerate,2]{label=\theenumi.\arabic*}
\setlist[enumerate,3]{label=\theenumii.\arabic*}

\begin{document}


\SetAuthorBlock{Saeid Bayat\CorrespondingAuthor}{Department of Industrial and Enterprise Systems Engineering,\\
   University of Illinois at Urbana-Champaign,\\
   Urbana, IL, USA \\
   email: bayat2@illinois.edu} 


\SetAuthorBlock{Nastaran Shahmansouri}{Autodesk Research,\\
   661 University Ave,\\
   Toronto, Canada \\
   email: nastaran.shahmansouri@autodesk.com} 
\SetAuthorBlock{Satya RT Peddada}{Department of Industrial and Enterprise Systems Engineering,\\
   University of Illinois at Urbana-Champaign,\\
   Urbana, IL, USA \\
   email: speddad2@illinois.edu} 
\SetAuthorBlock{Alex Tessier}{Autodesk Research,\\
   661 University Ave,\\
   Toronto, Canada \\
   email: alex.tessier@autodesk.com} 
\SetAuthorBlock{Adrian Butscher}{Autodesk Research,\\
   661 University Ave,\\
   Toronto, Canada \\
   email: adrian.butscher@autodesk.com} 
\SetAuthorBlock{James T Allison}{Department of Industrial and Enterprise Systems Engineering,\\
   University of Illinois at Urbana-Champaign,\\
   Urbana, IL, USA \\
   email: jtalliso@illinois.edu} 
   
\title{Extracting Design Knowledge from Optimization Data: Enhancing Engineering Design in Fluid Based Thermal Management Systems}

\keywords{Thermal Management System Design, Design Synthesis, Optimization, Optimal Flow Control, Knowledge Extraction, Graph Modeling}

\begin{abstract}
 As mechanical systems become more complex and technological advances accelerate, the traditional reliance on heritage designs for engineering endeavors is being diminished in its effectiveness. Considering the dynamic nature of the design industry where new challenges are continually emerging, alternative sources of knowledge need to be sought to guide future design efforts. One promising avenue lies in the analysis of design optimization data, which has the potential to offer valuable insights and overcome the limitations of heritage designs. This paper presents a step toward extracting knowledge from optimization data in multi-split fluid-based thermal management systems using different classification machine learning methods, so that designers can use it to guide decisions in future design efforts. This approach offers several advantages over traditional design heritage methods, including applicability in cases where there is no design heritage and the ability to derive optimal designs. We showcase our framework through four case studies with varying levels of complexity. These studies demonstrate its effectiveness in enhancing the design of complex thermal management systems. Our results show that the knowledge extracted from the configuration design optimization data provides a good basis for more general design of complex thermal management systems. It is shown that the objective value of the estimated optimal configuration closely approximates the true optimal configuration with less than 1 percent error, achieved using basic features based on the system heat loads without involving the corresponding optimal open loop control (OLOC) features. This eliminates the need to solve the OLOC problem, leading to reduced computation costs.
\end{abstract}

\date{Version \versionno, \today}

\maketitle 


\section{Introduction}
\label{Sec: Introduction}
Fluid-based thermal management systems play a critical role in various industries and applications, highlighting their significant importance \cite{buidin2021battery,bayat2023multisplit}. They ensure optimal performance, reliability, and longevity of components and equipment by regulating and controlling their temperature. Fluid-based thermal management systems are especially vital in high-power electronic devices, such as data centers, electric vehicles, and aerospace systems, where efficient heat dissipation is essential to prevent overheating and potential failures. These systems ensure the safe operation of electronic components by effectively managing heat through fluid circulation and heat exchange mechanisms \cite{mathew2022review}. This allows heat to be transferred from hot-spots to areas where dissipation is more efficient, such as heat sinks, maintaining desired operating temperatures and improving overall system performance. The significance of fluid-based thermal management systems extends beyond the realm of electronics. They are also essential in industrial processes, power generation, and environmental control systems \cite{liu2021review}. In these applications, efficient heat transfer and thermal regulation are crucial for maintaining optimal operational conditions, maximizing energy efficiency, and reducing environmental impact. In addition, as technology advances and systems become more powerful and compact, thermal management becomes increasingly important. Heat dissipation challenges escalate with higher power densities, making fluid-based thermal management systems indispensable for maintaining system reliability, performance, and safety \cite{laloya2015heat}.

Thermal management systems consist of various components. While enhancing each component individually can contribute to an improved design \cite{shah2005exergy}, it may not achieve optimality as this approach fails to consider the synergy between different components \cite{aloui2003handbook}. Conversely, when the configuration design of the entire system is conducted holistically, it has the potential to yield optimal results \cite{jafari2018thermal,bayat2023multisplit,SaeidBayat-Vehicle,bayat2023nested}.

Solving a particular configuration design problem is not the only use of configuration optimization methods. Rich sets of design data can also be produced that quantify desirable design characteristics across varying system needs and purposes. Such design data can augment historical design data (descriptions of engineering systems designed in the past and their resulting performance) to expand the set of information from which humans can derive generalizable engineering design knowledge. This is especially valuable when working to create unprecedented systems without design heritage \cite{Belm2018FromDT}, and when working to break free of the incremental progress that is associated with basing new designs on existing ones. In recent years, engineering design researchers have utilized data science techniques to extract design knowledge from past design data \cite{fuge2014machine}. However, these descriptive approaches have limitations as they require existing design heritage. Furthermore, these approaches are constrained by fixed data sets, lacking the ability to create new data sets to improve the quality of knowledge generation. On the other hand, design optimization data may reveal new patterns that lead us in new non-obvious directions. Knowledge extraction can be utilized to identify these patterns.

Several automated methods for extracting knowledge from design data have been explored. These methods encompass various forms of design knowledge and algorithmic approaches to facilitate efficient knowledge extraction. For example, Fuge et al.~\cite{fuge2014machine} demonstrated a machine learning approach for recommending appropriate design methods based on historical design data. Holzinger~\cite{holzinger2019introduction} articulates how artificial Neural Networks (NN) may be used for knowledge extraction. While NNs are useful for generalization, their complexity makes it difficult to extract knowledge through its structure \cite{MacDonald2020ExplainingNN, Liu2018ImprovingTI}.%

For the results to be human interpretable, complexity must be limited. For example, it might be helpful for engineers to gain a general sense of how an optimal design solution behaves as design scenarios change (e.g., loading conditions, design goals, design requirements). This knowledge is similar to the intuition of an experienced engineer who understands how to make suitable design adjustments to accommodate changing design needs based on the success and failure of previous experiences. Engineers who gain insight from knowledge extracted from data would then have a similar advantage to an engineer with deep intuition from experience. That said, knowledge from experience and insights from design optimization data do have some fundamental differences. The knowledge extraction process involves defining an optimization problem and solving it many times with different parameters. Following that, we extract knowledge about the relationship between the parameters and the results. The knowledge extraction from optimization data has two key advantages over the design heritage: 1) it is applicable in cases where there is no design heritage, and 2) the design is optimal, while the design derived from design heritage may not be optimal. 

In this article, we made an engineering system configuration design optimization framework for the investigation of thermal management systems. This framework is leveraged to generate design optimization data and perform design knowledge extraction. Various knowledge extraction algorithms are investigated to help build a foundation for designing a specific class of thermal management systems. 

The following are the main contributions of the work presented in this article:
\begin{enumerate}
    \item Developing interpretable knowledge based on optimization data that can be used to guide general design processes for a class of thermal management systems. This knowledge aids in estimating the optimal configuration of the fluid-based multi-split system, considering the given heat loads.
    \item Delving into a comprehensive analysis of the optimization data to aid the interpretation of the acquired knowledge.
    \item Leveraging the acquired knowledge to engineer novel and more intricate heat management systems.
\end{enumerate}

This article continues as follows: In Section \ref{Sec: System Description and Modeling}, we discuss the thermal management system architectures studied in this work and the corresponding Open Loop Optimal Control (OLOC) problem. Section~\ref{Sec: Method} explains the knowledge extraction method utilizing optimization data for single-split cases involving 3 and 4 Cold Plate Heat Exchangers (CPHXs), as well as multi-split cases with 3 CPHXs. Section \ref{Results} showcases four case studies aimed at evaluating the knowledge extracted from the previous section and its applicability in designing more complex heat management systems. The results demonstrate that the extracted knowledge can be effectively generalized to enhance the design of complex systems. In Section \ref{Limitations}, the limitations of this work are presented, along with potential suggestions for future research.  Section \ref{Conclusion} concludes with a summary of the design methodology and guidelines for thermal management system design.

\section{System Description and Modeling}
\label{Sec: System Description and Modeling}

\begin{figure}[ht!]
    \centering
    \includegraphics[width=1.0\linewidth]{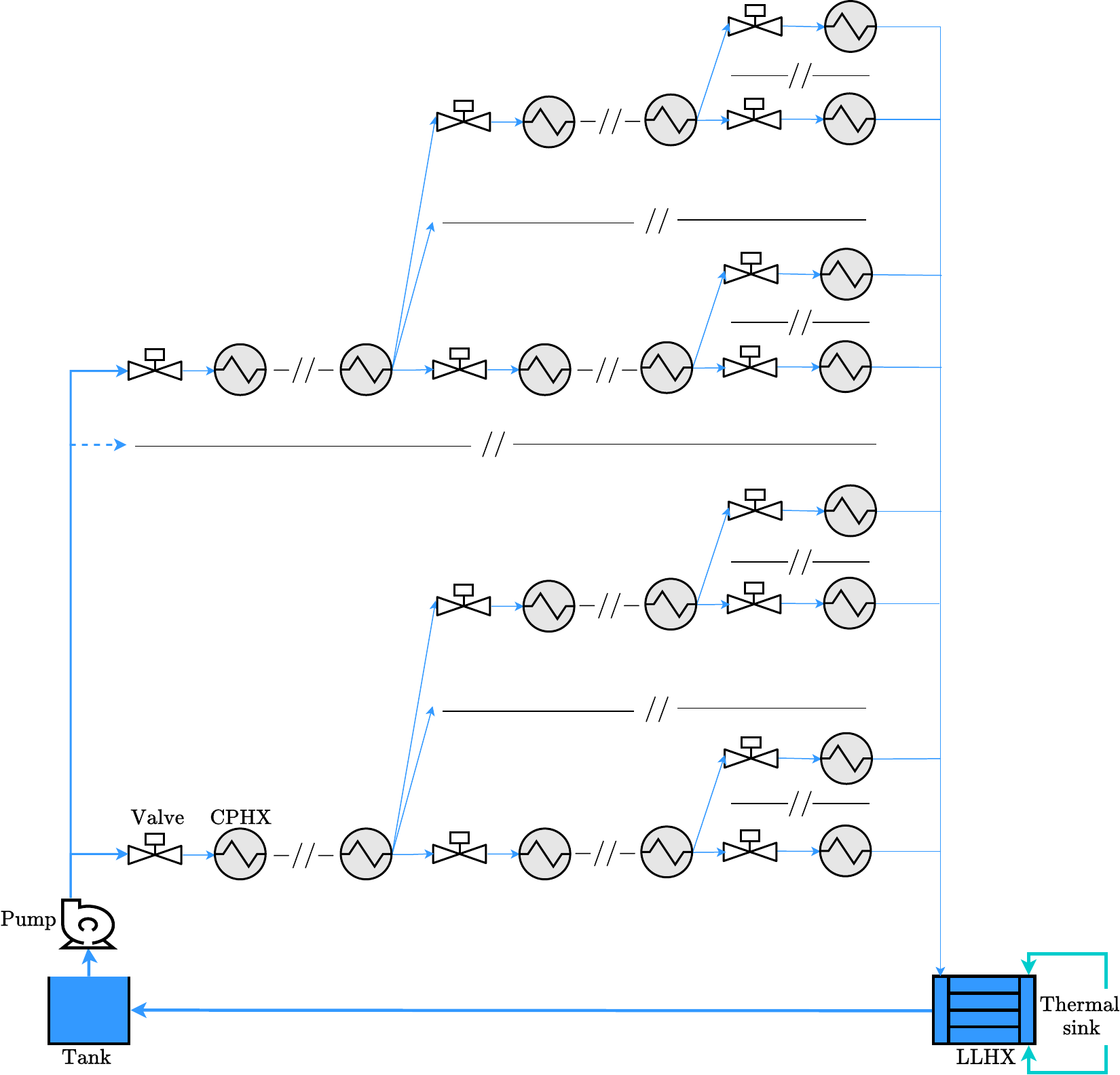}
    \caption{Class of problems considered in this paper. The systems include a tank, a pump, valve(s), CPHXs in parallel and series, a liquid-liquid heat exchanger (LLHX), and a sink.}
    \label{Fig:General_system}
\end{figure}

Figure~\ref{Fig:General_system} illustrates the thermal management system examined in this article, which aims to regulate the temperature of heat generating devices on CPHXs through a coolant flow. The system employs a multi-split configuration approach, as investigated in Ref.~\cite{bayat2023multisplit}, unlike previous single-split studies \cite{peddada2019optimal}, expanding the search space for improved performance. An algorithm is developed to generate multi-split architectures, allowing splits at the pump or CPHX locations. Optimal control problems are then defined for each configuration, seeking optimal flow rate trajectories to maximize system performance while considering temperature constraints. The models incorporate advection, convection, and bi-directional advection. The performance measure focuses on maximizing device operating time to maximize the thermal endurance, denoted as $t_{\mathrm{end}}$, while ensuring that temperature and mass flow rate limits are met.

The general code structure is depicted in Algorithm~\ref{Alg: Code Structure}. In this context, the \textit{data} refers to the positions of CPHX in [$x,y,z$] coordinates. The parameter \textit{numLevels} determines the depth at which the graph is analyzed to create junctions. For example, if \textit{numLevels} is set to 1, junctions can only be added to branches connected to the tank. Multiple graphs are generated based on the provided \textit{Data} and \textit{numLevels}, each having distinct configurations. These configurations can be parallel, series, or a combination of both. Within a set of $N$ graphs, a specific graph is indicated by the variable \textit{configNum}. The variable \textit{distrb} represents the heat load at each node, which is used as a basis for generating the graphs. Based on the provided data, two types of graphs are generated: the \textit{Base Graph} and the \textit{Physics Graph}. The \textit{Base Graph} represents the topology of the thermal system, or how the tanks and the CPHXs are connected, while the \textit{Physics Graph} includes additional nodes such as walls and sinks, and incorporates the dynamics between these nodes.

Subsequently, the Optimal Control (OLOC) problem is defined. In this problem, the control variable is the derivative of the flow rate for each independent branch, and the state variables include the node temperatures and the flow rate of independent branches. Constraints are imposed to ensure that the input flow rate of each branch is equal to the output flow rate of all its sub-branches. Additionally, the sum of flow rates in all end branches must equal the pump flow rate ($\dot{m}_p$). These constraints are used to determine the flow rate of dependent branches ($\dot{\bm{m}}_{\mathrm{dp}}$).

Furthermore, initial temperatures are defined for the wall (w), fluid (f), and sink (l) nodes. Path constraints are established to ensure that temperature nodes satisfy certain upper bounds throughout the entire time horizon. Other path constraints are imposed to restrict the flow rates to be equal or less than the pump flow rate and to limit the maximum value of the control signal (derivative of the flow rate) for independent branches.

The convergence of the algorithm is achieved when at least one temperature reaches the upper bound; the time at which this occurs is reported as the objective function value. Table \ref{tab: Problem parameters} shows the parameters used for the studies in this article. In this paper, the OLOC problem is solved by Dymos~\cite{falck2021dymos}, an open-source Python program, using direct method (discretize and optimize) for solving the optimization problem; this method is suitable for solving complex problems \cite{gill2005snopt,biegler2009large}. In the future, Model Predictive Control (MPC) \cite{bayat2023ss,bayat2023lgrmpc} will be investigated for its closed-loop capabilities,

\begin{algorithm}[ht!]
  \caption{Graph Generation, Modeling, and Optimization Framework}\label{Alg: Code Structure}
  \footnotesize
  \begin{algorithmic}[]
  \State Define CPHX-Position in [x,y,z] coordinates
  \State \hspace{0.5cm} Ex: $Data=[[2,0,0],[3,1,0],[10,10,0],[15,10,0]]$
  \vspace{0.2cm}
  \State Define Heat Load $\&$ Graph Config
  \State \hspace{0.5cm} Ex: $numLevels=1$
  \State \hspace{0.9cm} $configNum=0$
  \State \hspace{0.9cm} $distrb=[5,5,5,5,5,5]$ kW
  \State \hspace{0.9cm} $p=\textrm{Thermal Properties}$
  \vspace{0.2cm}
  \State Generate base graph and then physics graph
  \State \hspace{0.5cm} 
  \begin{center}
  \includegraphics[width=0.1\textwidth]{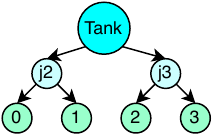}\hspace{0.5cm}
  \includegraphics[width=0.3\textwidth]{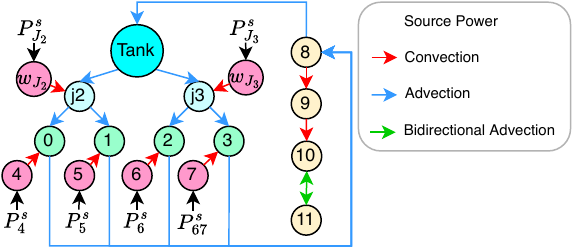}
  \end{center}
  \vspace{0.2cm}
    \State Run OLOC
    \State \hspace{0.5cm} Define states: All temperatures and flow rate of independent branches:
    \State \hspace{3.0cm} $\bm{\xi}=[\bm{T},\dot{\bm{m}}_{\mathrm{indp}}]$
    \vspace{0.1cm}
    \State \hspace{0.5cm} Define controls: Derivative of flow rate of independent branches:
    \State \hspace{3.0cm} $\bm{u}=[\ddot{\bm{m}}_{\mathrm{indp}}]$
    \vspace{0.1cm}
    \State \hspace{0.5cm} Define dependent flow rate of independent branches
    \State \hspace{3.0cm} $\dot{\bm{m}}_{\mathrm{dp}}=\mathbf{M} \times \dot{\bm{m}}_{\mathrm{indp}}$
    \vspace{0.1cm}
    \State \hspace{0.5cm} Define initial condition, bound, dynamics, and path constraint
    \State \hspace{1.0cm} $\bm{T}_w(0)=\bm{T}_{w,0}$ \hspace{1.0cm} $\bm{T}_f(0)=\bm{T}_{f,0}$ \hspace{1.0cm} $\bm{T}_l(0)=\bm{T}_{l,0}$
    \vspace{0.1cm}
    \State \hspace{1.0cm} $\bm{T}_w(t)\le \bm{T}_{w,\mathrm{max}}$ \hspace{0.75cm} $\bm{T}_f(t)\le \bm{T}_{f,\mathrm{max}}$ \hspace{0.75cm} $\bm{T}_l(0) \le \bm{T}_{l,\mathrm{max}}$
    \vspace{0.1cm}
    \State \hspace{1.0cm} $0 \le \dot{\bm{m}}_{\mathrm{dp}} \le \dot{\bm{m}}_{\mathrm{p}} $ \hspace{0.85cm} $0 \le \dot{\bm{m}}_{\mathrm{indp}} \le \dot{\bm{m}}_{\mathrm{p}} $ \hspace{0.75cm} $| \bm{u} | \le \ddot{\bm{m}}_{f,\mathrm{max}}$
    \vspace{0.1cm}
    \State \hspace{1.0cm} $\dot{\bm{T}}=\bm{A} \begin{bmatrix} \bm{T} \\ T^t \end{bmatrix} +\bm{B}_1 \left( diag \left( \bm{Z} \begin{bmatrix} \dot{m}_{\mathrm{p}} \\ \dot{\bm{m}}_f \\ \dot{m}_t\end{bmatrix} \right) \right) \bm{B}_2 \begin{bmatrix} \bm{T} \\ T^t \end{bmatrix} + \bm{C}^{-1} D P^s$
    \vspace{0.2cm}
    \State \hspace{0.5cm} Define objective
     \State \hspace{3.3cm} $\max t_{\mathrm{end}}$
  \end{algorithmic}
  \end{algorithm} 

\begin{table}[ht!]
\small
\centering
\caption{Parameters used in the physics modeling of the thermal systems.}
\label{tab: Problem parameters}
\scalebox{0.8}{
\begin{tabular}{rlrl}
\toprule
Parameter  & Value &   Parameter  & Value    \\ \hline
CPHX wall mass                      & 1.15 kg    &
LLHX wall mass                      & 1.2 kg     \\
Tank fluid mass                     & 2.01 kg    &
Thermal sink temperature $T^t$         & $15^o$ C       \\
initial temperature , $T_{l,0}$ & $15^o$ C       &
wall temperature, $T_{w,0}$  & $20^o$ C       \\
fluid temperatue , $T_{f,0}$  & $20^o$ C       &
Thermal sink mass flow rate, $\dot{m}_t$ & 0.2 kg/s   \\
Pupmp mass flow rate $\dot{m}_p$          & 0.4 kg/s   &
Valve rate limit $\ddot{m}_{f,\mathrm{max}}$              & 0.05 kg/$\mathrm{s}^2$ \\ \bottomrule
\end{tabular}
}
\end{table}

\section{Training Procedure}
\label{Sec: Method}

There are two scenarios in designing a classifier: 1) designing a simple classifier where the result is interpretable by humans, but the accuracy is not necessarily very high, and 2) designing a complex classifier with more features with high accuracy where the result is difficult to interpret. This paper pursues the first scenario to get optimization data that is interpretable by humans or other optimization tools. We aim to use a limited number of features, without considering OLOC related variables, to estimate the optimal configuration. This approach allows designers to estimate which cases may yield better design solution, without solving the OLOC problem.

Algorithm~\ref{Training Procedure} gives an overview of the training procedure employed in this article. Initially, the system is provided with the population size, heat load range, and number of nodes (CPHXs). In the algorithm, $n_{\mathrm{conf}}$ represents the total number of configurations for graphs with $n_{\mathrm{nodes}}$. Based on the chosen sampling method, a population of heat loads (disturbances) is generated for each configuration. For each member of the population, the configuration with the maximum thermal endurance is identified using the code from Algorithm~\ref{Alg: Code Structure}. The best configuration is assigned as the label to that member of the population. Subsequently, new features are defined, and a classification algorithm is trained to estimate the best label based on these features. This training is conducted on the training data and later tested on the test data.

\begin{algorithm}[ht!]
\footnotesize
  \caption{Training Procedure}\label{Training Procedure}
  \begin{algorithmic}[1]
    \State $n_{\mathrm{pop}} \gets $ Population size
    \State $d_{\mathrm{range}} \gets $ Heat load range
    \State $n_{\mathrm{nodes}} \gets $ Number of nodes
    \State ${n_{\mathrm{conf}}} :$ Compute number of different configurations 
    \State $d_{n_{\mathrm{pop}} \times n_{\mathrm{nodes}}} : $ Generate a population of heat load samples
    \vspace{0.1cm}
    \State $L_{\mathrm{pop}} \gets \bm{0}$ (Initialization of best label for each heat load)
    \vspace{0.1cm}
    \For{$i $ in range($n_{\mathrm{pop}}$)} 
    \State $d_i \gets d_{n_{\mathrm{pop}} \times n_{\mathrm{nodes}}}[i,:]$
    \vspace{0.1cm}
    \State $J_{n_{\mathrm{conf}}} \gets $ Run the code given $d_i$ and $n_{\mathrm{nodes}}$ and get thermal endurance
    \vspace{0.1cm}
    \State $L_{\mathrm{pop}}[i] \gets $ $\operatorname*{argmax}_{i} J_{n_{\mathrm{conf}}}$ (Get the best label for that disturbance)
    \EndFor
    \State $n_{\mathrm{f}} \gets $ Number of features
    \vspace{0.1cm}
    \State $D_{n_{\mathrm{pop}} \times n_{\mathrm{f}}} \gets $ Generate features based on $d$
    \vspace{0.1cm}
    \State $n_{\mathrm{train}} \, , n_{\mathrm{test}}  \gets $ Set Number of training and test points among a total of $n_{\mathrm{pop}}$
    \vspace{0.1cm}
    \State $G(D_{n_{\mathrm{train}} \times n_{\mathrm{f}}}) : $ Train the classifier
    \State $\hat{L}_{n_{\mathrm{test}}} \gets G(D_{n_{\mathrm{test}} \times n_{\mathrm{f}}}) : $ Test the classifier on the test data
  \end{algorithmic}
\end{algorithm}

We present classifications of three different scenarios. Section \ref{sec: CPHx with 3 nodes- Single Split} investigates single split cases with 3 nodes, comprising a total of 13 different configurations. These cases are examined under 400 different disturbances. Various machine learning methods are employed for the data classification. In Section \ref{sec: CPHx with 3 nodes- Multi Split}, we explore multi-split cases with 3 nodes, encompassing 3 distinct classes. These cases are also analyzed using the same 400 disturbances as the ones used in the studies of Section \ref{sec: CPHx with 3 nodes- Single Split}. Additionally, in Section \ref{sec: CPHx with 4 nodes- Single}, we examine single split cases with 4 nodes, which consist of 73 different configurations. These cases are studied under 200 different disturbances. Finally, the utilization of the trained models for designing new systems is explained in Section \ref{Results}.

\subsection{Single-Split Graphs with 3 Nodes }
\label{sec: CPHx with 3 nodes- Single Split}

\begin{figure}[ht!]
    \centering
    \includegraphics[width=1.0\linewidth]{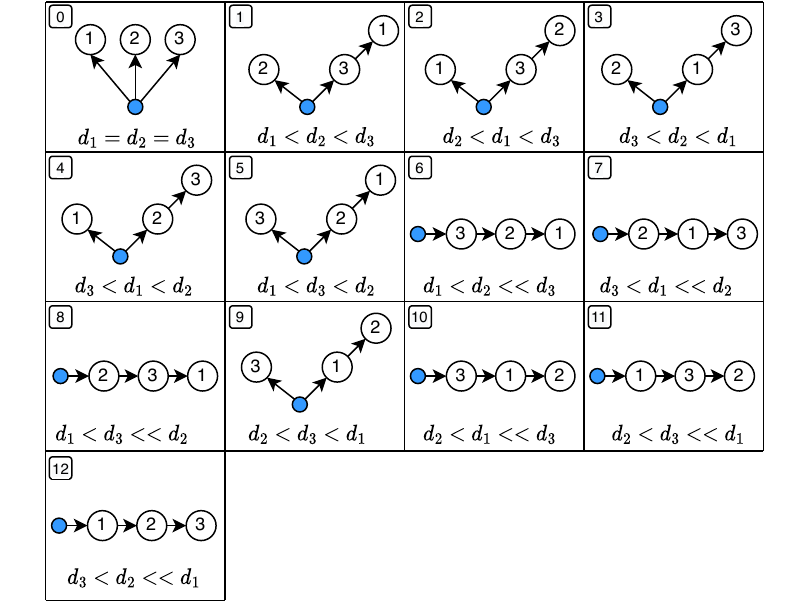}
    \caption{Single-split structures with 3 heat generating components. The heat load condition at which each configuration is optimal is also presented.}
    \label{fig:Single_3_configs}
\end{figure}

Graph consisting of three nodes are investigated in this section. The disturbance for each node (i.e. CPHX) is considered within the range of $d_i \in [4, 16] , \mathrm{kW}$, where $i , \in ,{1, 2, 3}$. To explore various disturbances, we employ Latin Hypercube Sampling (LHS) method, generating a population of 400 disturbances. Subsequently, we utilize the algorithm described in the previous section to solve the OLOC problem and obtain the optimal objective function value. Each disturbance corresponds to 13 different graphs with distinct configurations, as shown in Fig.~\ref{fig:Single_3_configs}. This figure showcases all 13 configurations and the conditions under which each configuration represents the optimal structure among all others. For instance, configuration 0 is optimal when all disturbances are close together, while configuration 12 is the optimal configuration when the first disturbance is significantly higher than the other disturbances. Throughout this section, the process of classification and the derivation of knowledge will be demonstrated.

In the classification process, to efficiently solve all 13 configurations simultaneously, parallel computation is employed. The values of the parameters in Algorithm~\ref{Training Procedure} used in this study are shown in Table ~\ref{tab:alg_params_3odes_single}. Initially, the features, denoted as $D$, are selected to be identical to the disturbances. Subsequently, the normalized value of disturbances are used to aid generalization.

\begin{table}[ht!]
\small
\centering
\caption{Values of parameters in Algorithm~\ref{Training Procedure} used in Sec.~\ref{sec: CPHx with 3 nodes- Single Split} for single split cases with 3 CPHXs.}
\label{tab:alg_params_3odes_single}       
\scalebox{1.0}{
\begin{tabular}{cccc}
\toprule
param     & value  & param        & value  \\ \hline
$n_{\mathrm{pop}}$   & 400 & $d_{\mathrm{range}}$ & $[4,16]$kW \\
$n_{\mathrm{nodes}}$ & 3  & ${n_{\mathrm{conf}}}$   & 13 \\
$n_{\mathrm{f}}$ & 3 & $D$       & $d_i$ or $d_i/\Sigma d_i$ \\
$n_{\mathrm{train}}$ & 350 & $n_{\mathrm{test}}$ &  50 \\
Sampling method   & LHS & & \\
\bottomrule
\end{tabular}
}
\end{table}

\begin{figure}[ht!]
    \centering
    \subcaptionbox{}{\includegraphics[width=0.49\linewidth]{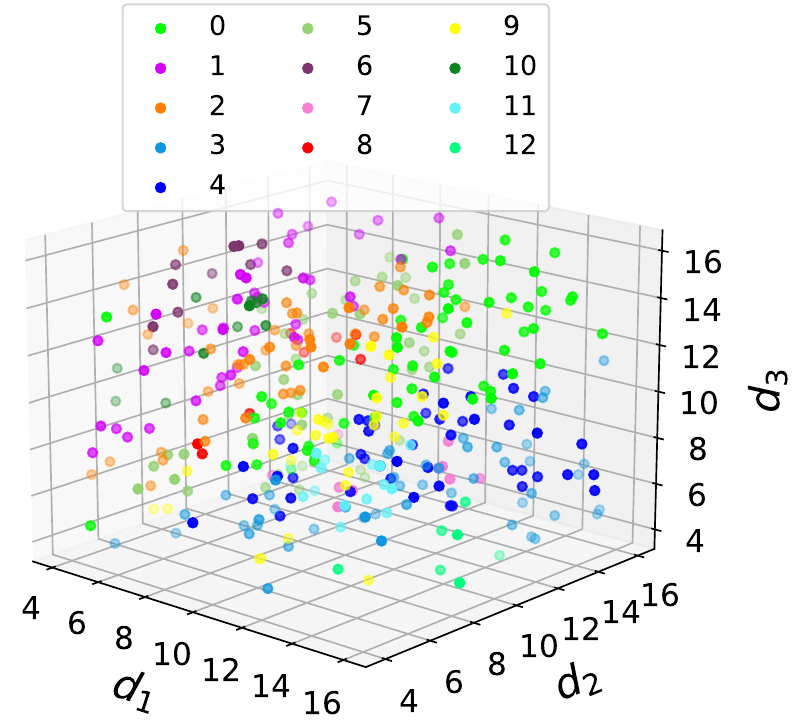}}
    \subcaptionbox{}{\includegraphics[width=0.49\linewidth]{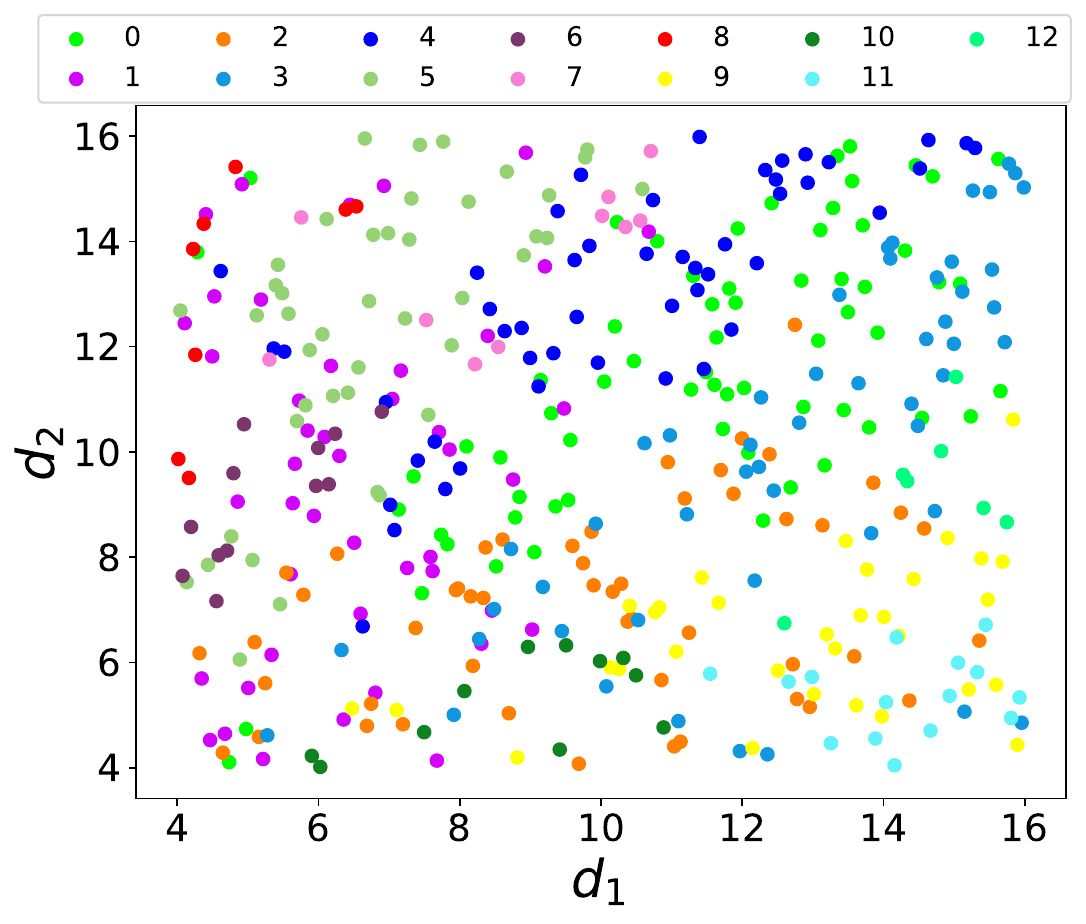}}
    \subcaptionbox{}{\includegraphics[width=0.49\linewidth]{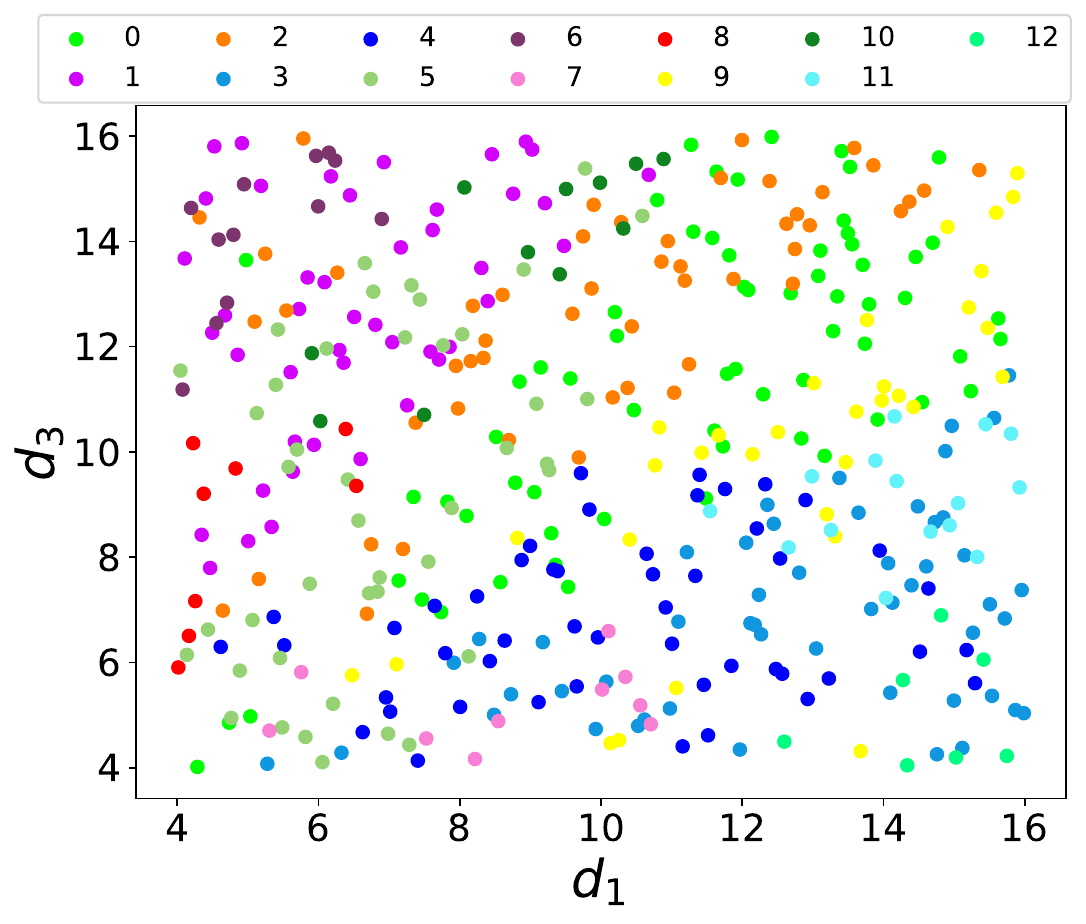}}
    \subcaptionbox{}{\includegraphics[width=0.49\linewidth]{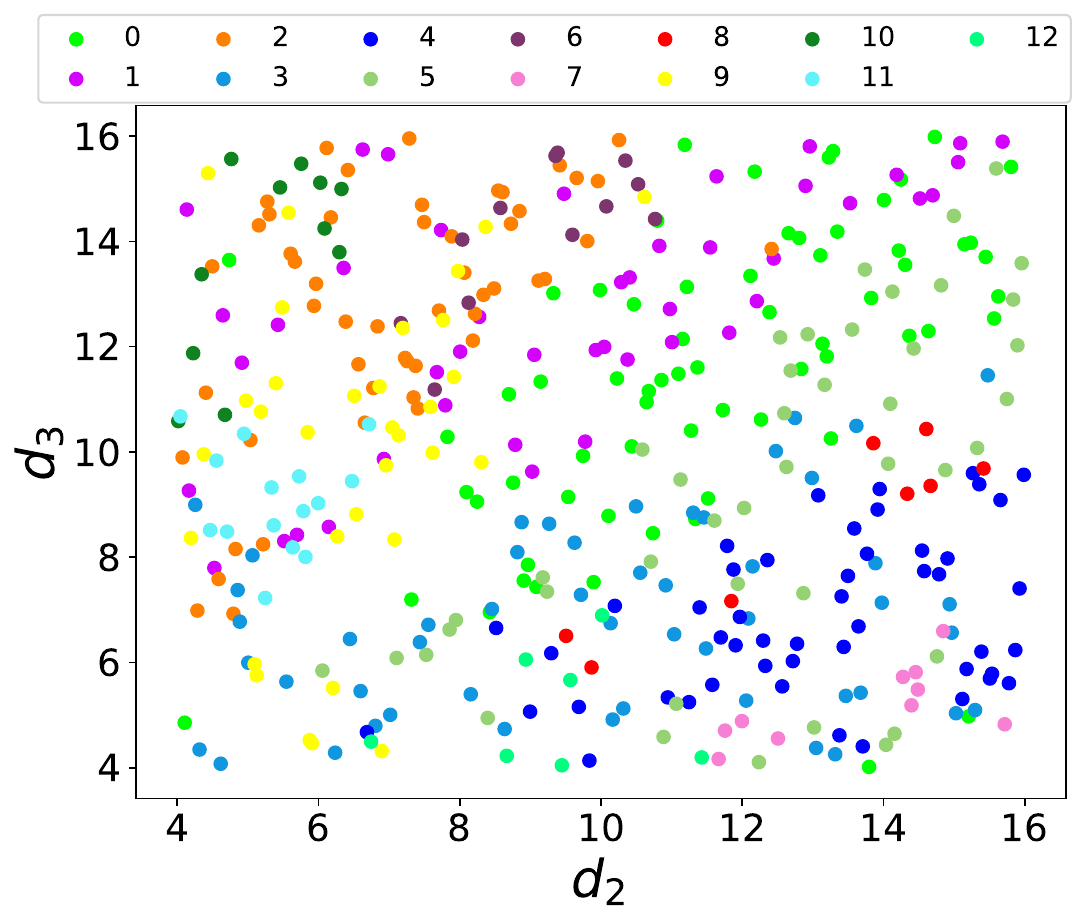}}
    \caption{Population obtained from hypercube sampling based on raw features for single-split cases with 3 CPHXs. All units are in kW.}
    \label{fig:scatter single3 Pop}
\end{figure}

Figure~\ref{fig:scatter single3 Pop} illustrates the generated population in 2D and 3D domains, where each axis represents a disturbance and the legend displays the corresponding labels for each class. As discussed previously, each disturbance results in 13 graphs with different configurations. The label assigned to a disturbance represents the configuration that yields the maximum objective function value among the 13 cases. For instance, dark-blue corresponds to a disturbance where the fourth configuration attains the highest objective function value.

Upon examining the data based on these raw features, it becomes apparent that the data points are not easily separable. To address this issue, we introduce three new features: $D_{i}=d_i/\Sigma_{i=1}^{3}d_i$ for i={1,2,3}. The utilization of dimensionless values facilitates generalization to different scenarios and helps comparing the cases based on relative disturbance values. For instance, even when two cases possess entirely distinct disturbances, they can be considered as representing the same scenario in the feature domain if their relative values remain consistent. Additionally, the third feature, $D_3$, is excluded from the feature set due to its linear relationship with $D_1$ and $D_2$ ($D_3=1-(D_1+D_2)$). This also helps reduce the number of feature. During training, it was also noticed that removing this feature does not negatively impact the accuracy of the model.

\begin{figure}[ht!]
    \centering
    {\includegraphics[width=1.0\linewidth]{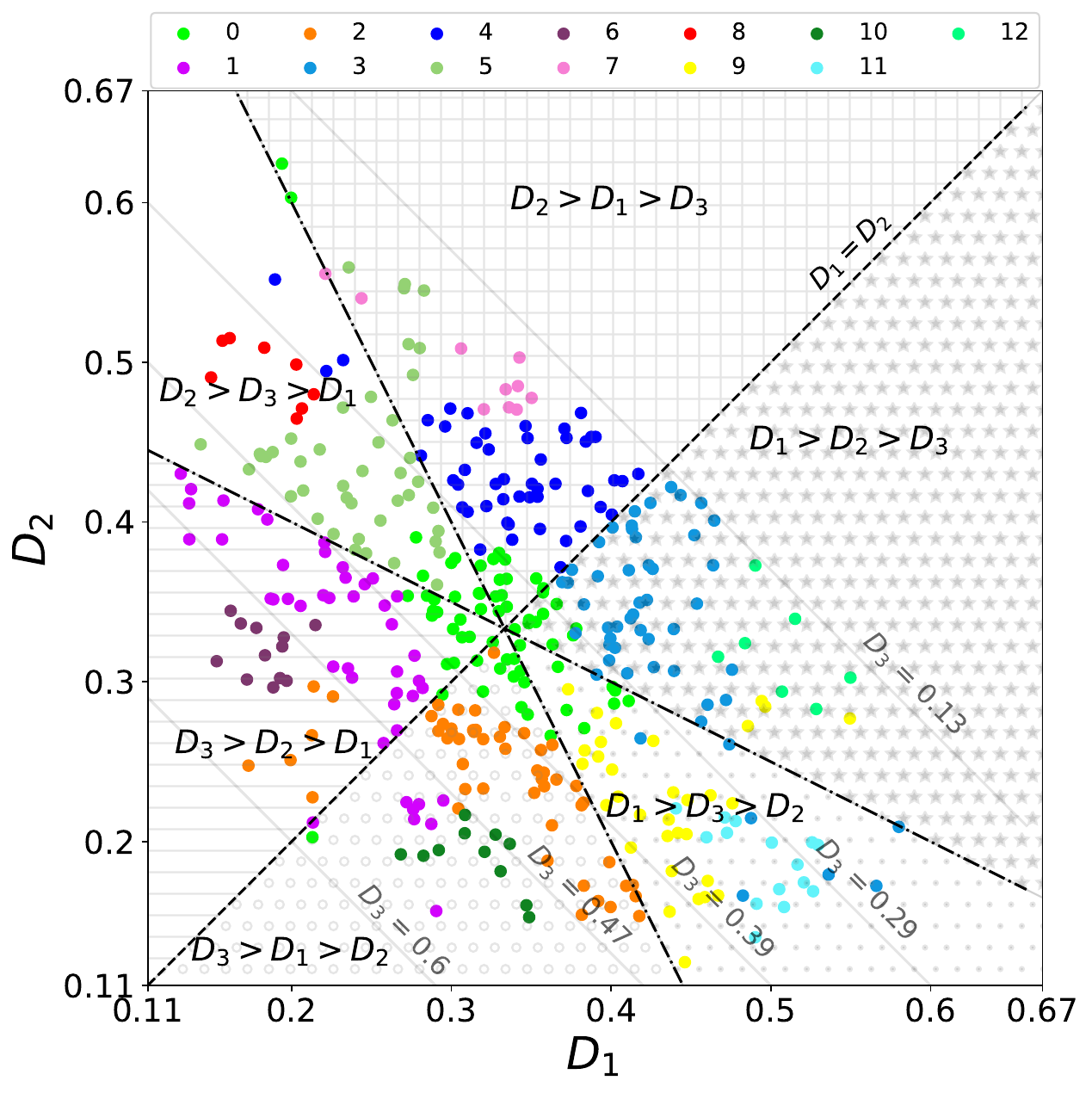}}
    \caption{Population obtained from hypercube sampling after feature selection defined as $D_1=d_1/\Sigma_{i=1}^3 d_i$, and $D_2=d_2/\Sigma_{i=1}^3 d_i$, for single--split cases.}
    \label{fig:scatter single3 Pop features}
\end{figure}

Figure~\ref{fig:scatter single3 Pop features} displays the population after applying the transformations based on these features. The x-axis represents $D_1$, the y-axis represents $D_2$, and each color corresponds to the best configuration among 13 different configurations within the feature space. The figure includes three guidelines that divide the feature space into six distinct sections, each represented by a different pattern. These sections illustrate varying relationships between the magnitudes of $D_1$, $D_2$, and $D_3$. For instance, in the center-top section, the following inequality holds: $D_2>D_1>D_3$. These six regions provide an approximate understanding of the conditions under which each graph (out of the 13 possible graphs) represents the optimal configuration. For example, a large population of class 4 resides in the top-center region, indicating that this configuration is optimal when the relationship between features is expressed as $D_2>D_1>D_3$. Similarly, class 0 is situated in the center, suggesting that configuration 0 is the optimal choice when all features have the same magnitude, i.e., $D_1\approx D_2\approx D_3$.

Figure~\ref{fig:scatter single3 Pop features} also includes additional lines indicating the magnitude of $D_3$, with all points on each line sharing the same $D_3$ value. These lines prove beneficial in enhancing the accuracy of estimating the optimal configuration within each region. This is crucial since, in each region, multiple configurations coexist. For example, in the lower-left region, three configurations dominate: 0, 2, and 10. The value of $D_3$ serves to discriminate between these configurations effectively. For instance, within this region, when $D_3<0.39$, configuration 0 is the most suitable; when $0.39<D_3<0.47$, configuration 2 performs best; and when $0.47<D_3<0.6$, configuration 10 becomes optimal. It's worth noting that $D_3$ is not defined as a feature for classification due to its linear relationship with $D_1$ and $D_2$. Consequently, expressing $D_3<0.39$ is equivalent to stating $D_1+D_2>0.61$. During training, only $D_1$ and $D_2$ are considered as features, indirectly indicating the value of $D_3$. As evident in the figure, the data points in the feature space become separable, thus significantly aiding the classification process.

\begin{figure}[ht!]
    \centering
    {\includegraphics[width=1.0\linewidth]{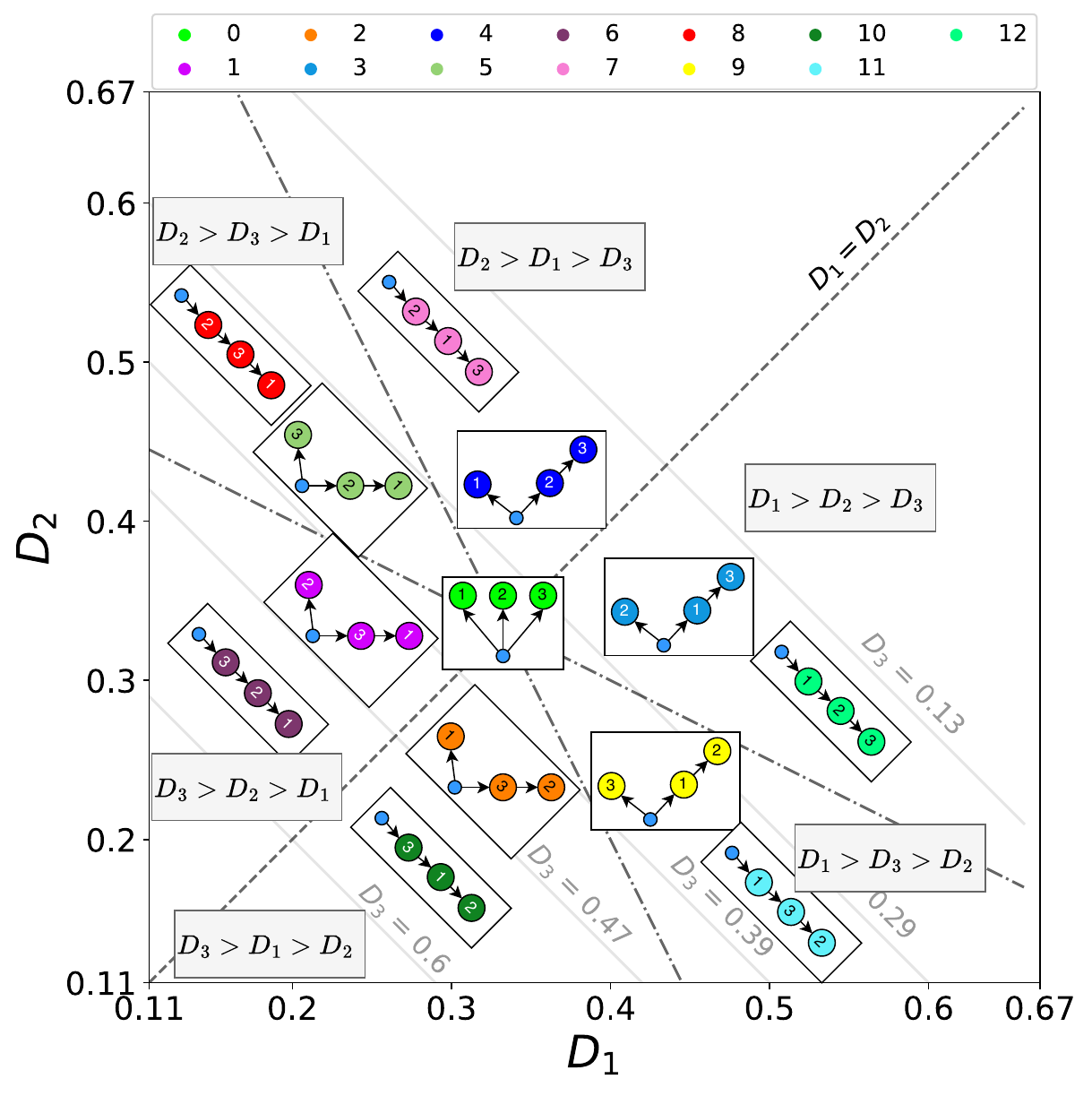}}
    \caption{Location of Optimal Configurations in the Feature Domain for Single Split Cases with 3 Nodes (Fig.\ref{fig:Single_3_configs}). The feature selection defined as $D_1=d_1/\Sigma_{i=1}^3 d_i$, and $D_2=d_2/\Sigma_{i=1}^3 d_i$}
    \label{fig:singke_3_null}
\end{figure}

The 13 graphs previously shown in Fig.~\ref{fig:Single_3_configs} are now plotted in the feature domain in Fig.~\ref{fig:singke_3_null}. The positioning of each graph in this plot is determined based on Fig.~\ref{fig:scatter single3 Pop features}, where the concentration of each class in a region influences its location. For example, class 0 was predominantly concentrated in the center, where $D_1 \approx D_2 \approx D_3$, and in this plot, graph 0 is appropriately placed at the center where this condition holds. As observed, graph 0 represents the configuration where all branches are parallel, positioned right at the center. In the six different regions surrounding class 0, we find other configurations with two branches, featuring a single node in one branch and two nodes in the other branch. Additionally, six different cases with all three nodes placed in series are positioned farther away from the center. The placement of these graphs in Fig.~\ref{fig:singke_3_null} is directly influenced by the concentration of each class in specific regions within the feature space, accurately representing the different optimal configurations based on the corresponding relationships between $D_1$, $D_2$, and $D_3$.

The placement of different graphs based on the feature space in Fig.~\ref{fig:singke_3_null} seems reasonable. For instance, in the top center region where $D_2>D_1>D_3$, classes 4 and 7 are considered optimal, and in both cases, node 2, which experiences the maximum disturbance, is located closest to the tank. Additionally, for class 4, the difference between $D_2$ and $D_1$ is not substantial. However, as we move to the top left (class 7), $D_2$ significantly surpasses $D_1$. In this situation, with $D_2>>D_1>D_3$, node 2 must receive the maximum fluid from the pump, which is achieved by placing all CPHXs in series. Conversely, for class 4, where $D_2$ is not much higher than the other disturbances, node 2 does not require maximum fluid. Solving the OLOC problem for this condition where $D_2$ is not much higher than $D_1$ revealed that this configuration (4) yields a better objective value function than case 7.

Moreover, in case 4, we observe that node 3 is positioned after node 2, which is reasonable as node 2 experiences higher disturbances and must be located close to the pump to receive cool fluid. The fluid reaching node 3 has already absorbed some heat from node 2. Similarly, in case 0, where features are close ($D_1 \approx D_2 \approx D_3$), the optimal solution involves placing all branches in parallel. This configuration grants the system maximum control authority since flow rates in each branch can be altered through valves. While we have discussed specific regions in Fig.~\ref{fig:singke_3_null}, the same concepts and similar explanations apply to different regions.

\begin{figure}[ht!]
    \centering
    \subcaptionbox{config 0}{\includegraphics[width=0.32\linewidth]{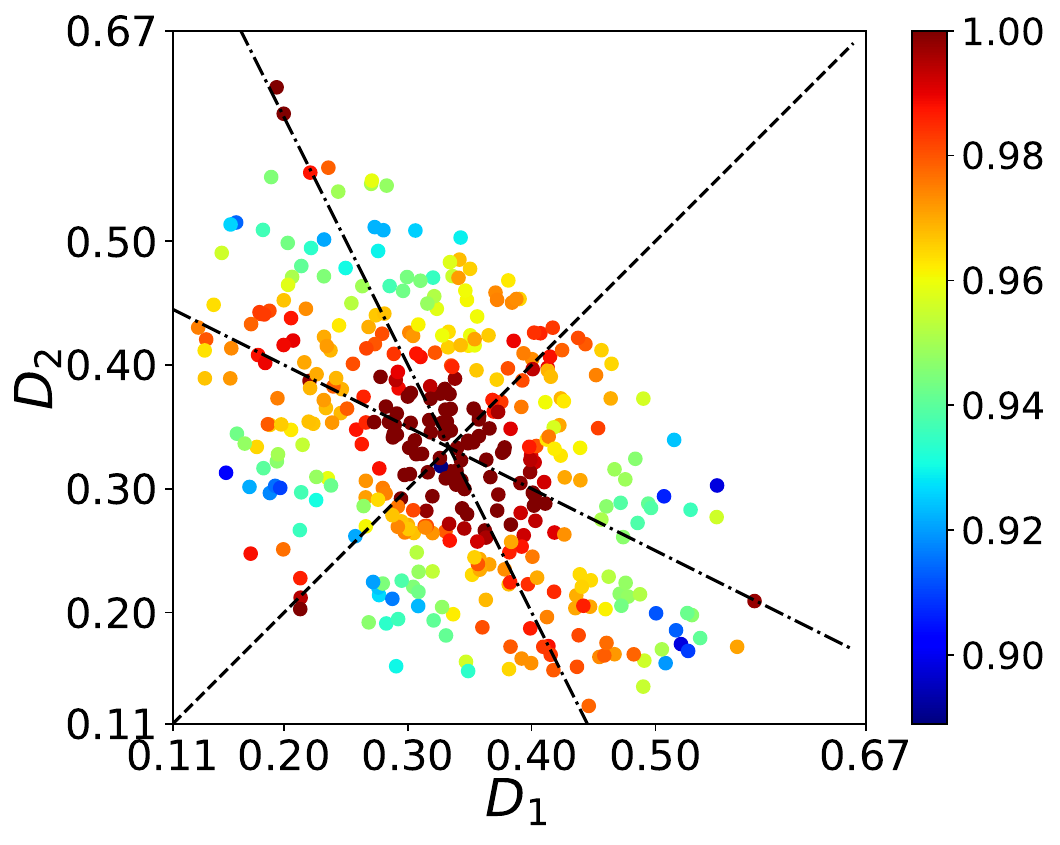}}
    \subcaptionbox{config 1}{\includegraphics[width=0.32\linewidth]{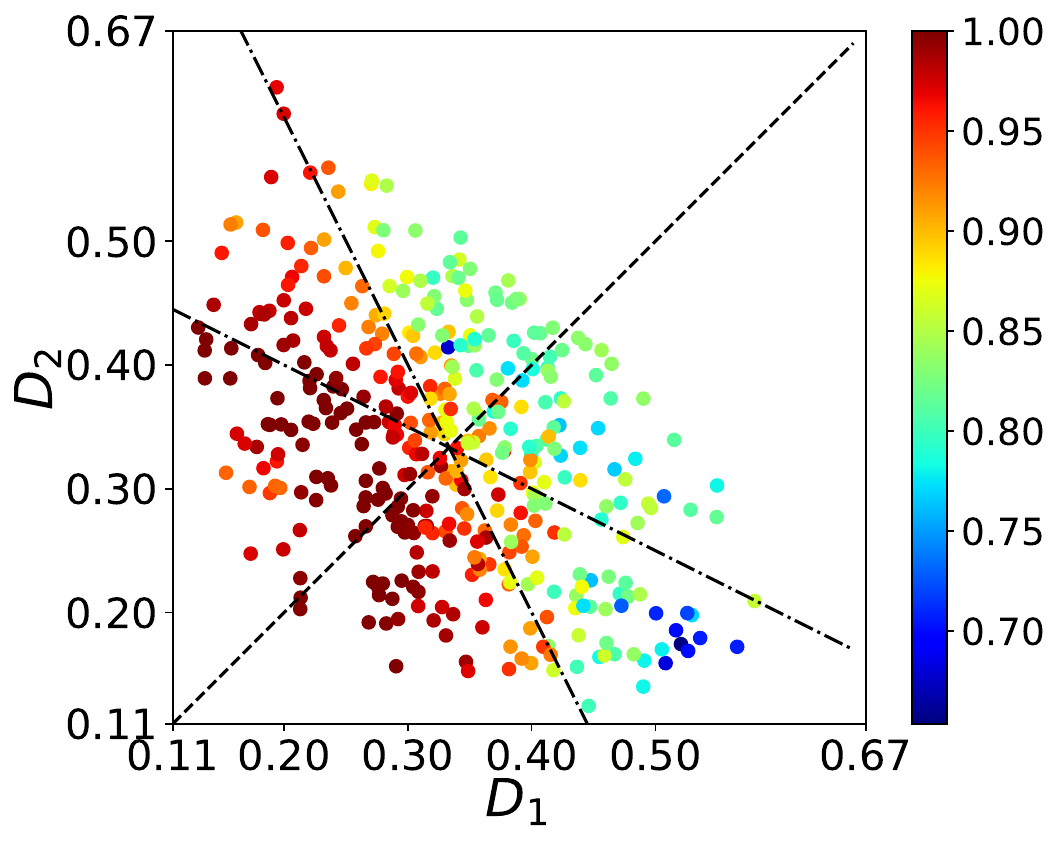}}
    \subcaptionbox{config 2}{\includegraphics[width=0.32\linewidth]{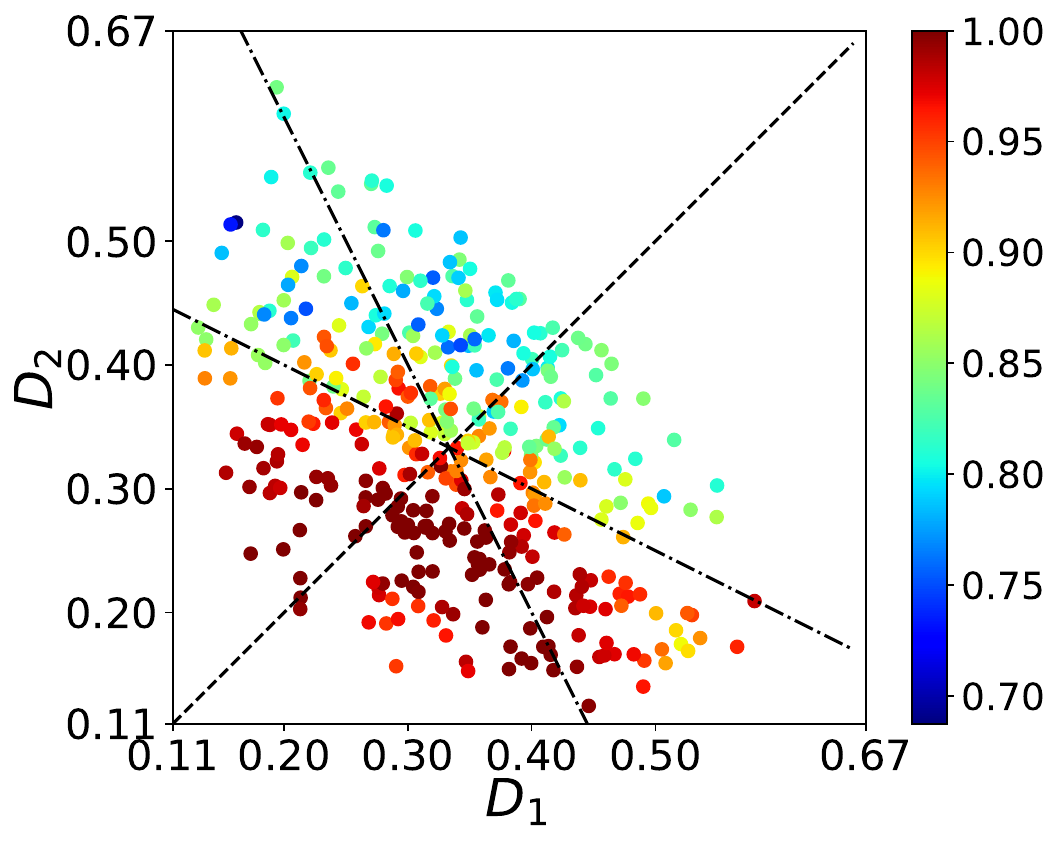}}
    \subcaptionbox{config 3}{\includegraphics[width=0.32\linewidth]{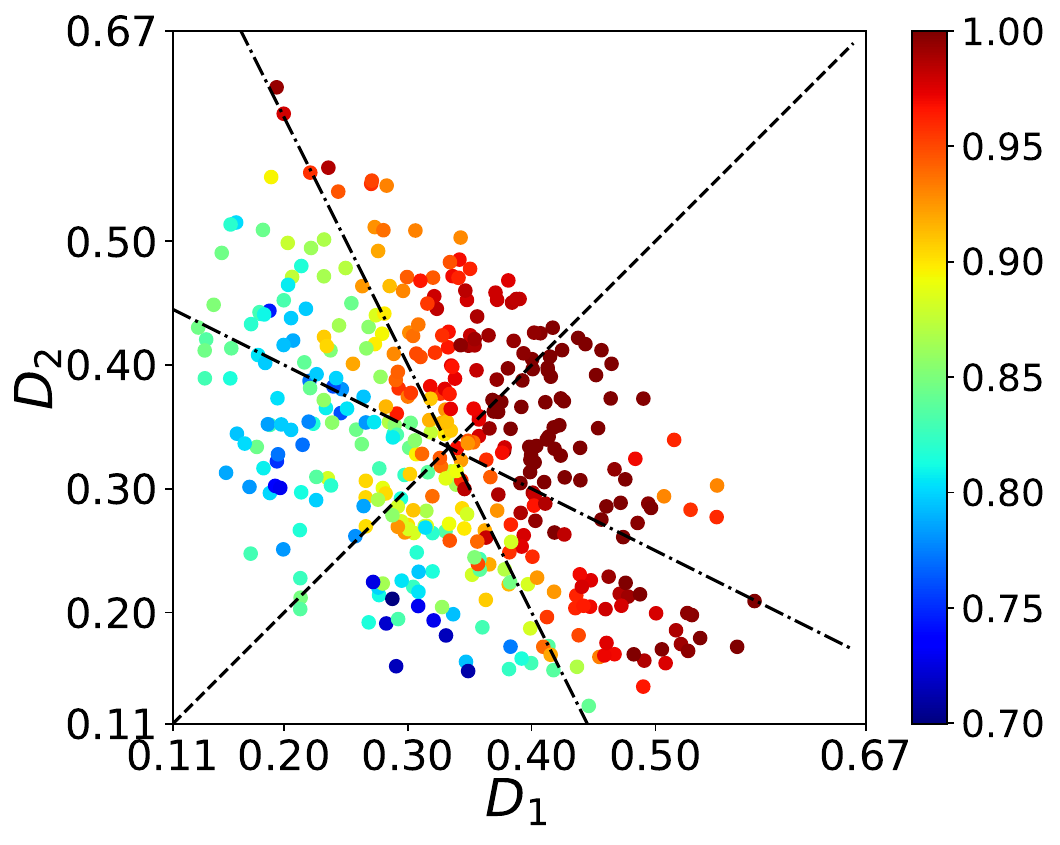}}
    \subcaptionbox{config 4}{\includegraphics[width=0.32\linewidth]{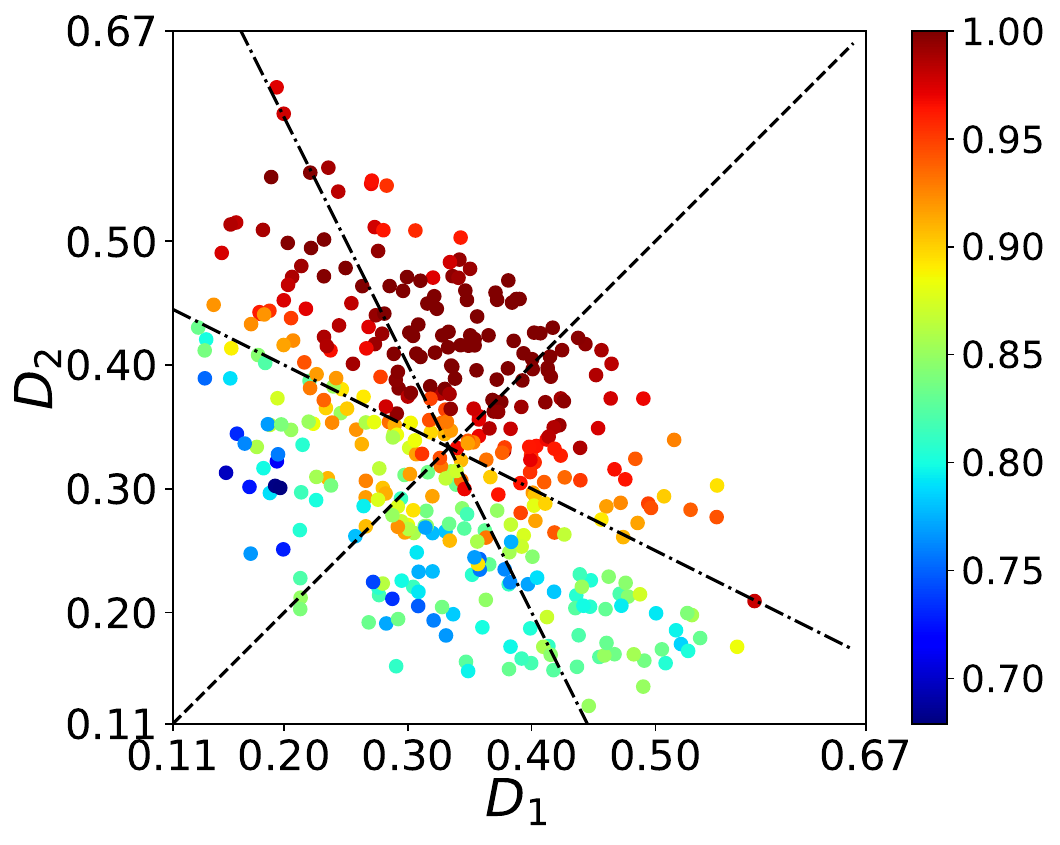}}
    \subcaptionbox{config 5}{\includegraphics[width=0.32\linewidth]{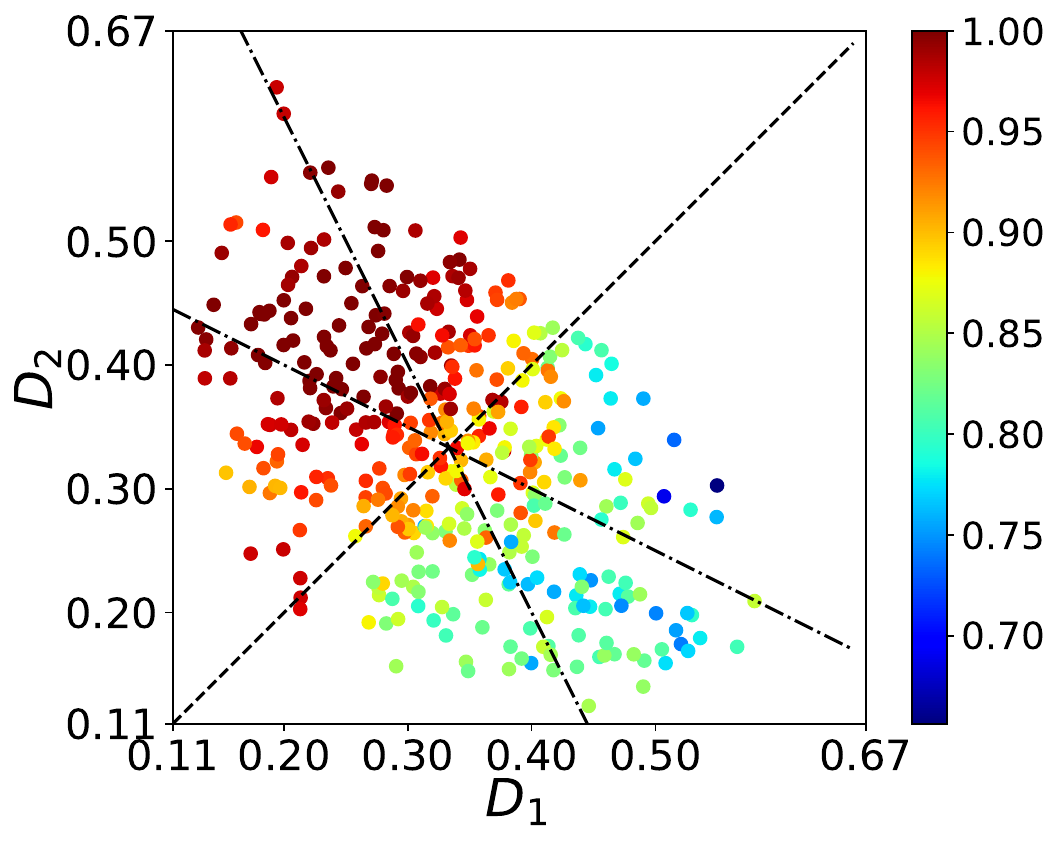}}
    \subcaptionbox{config 6}{\includegraphics[width=0.32\linewidth]{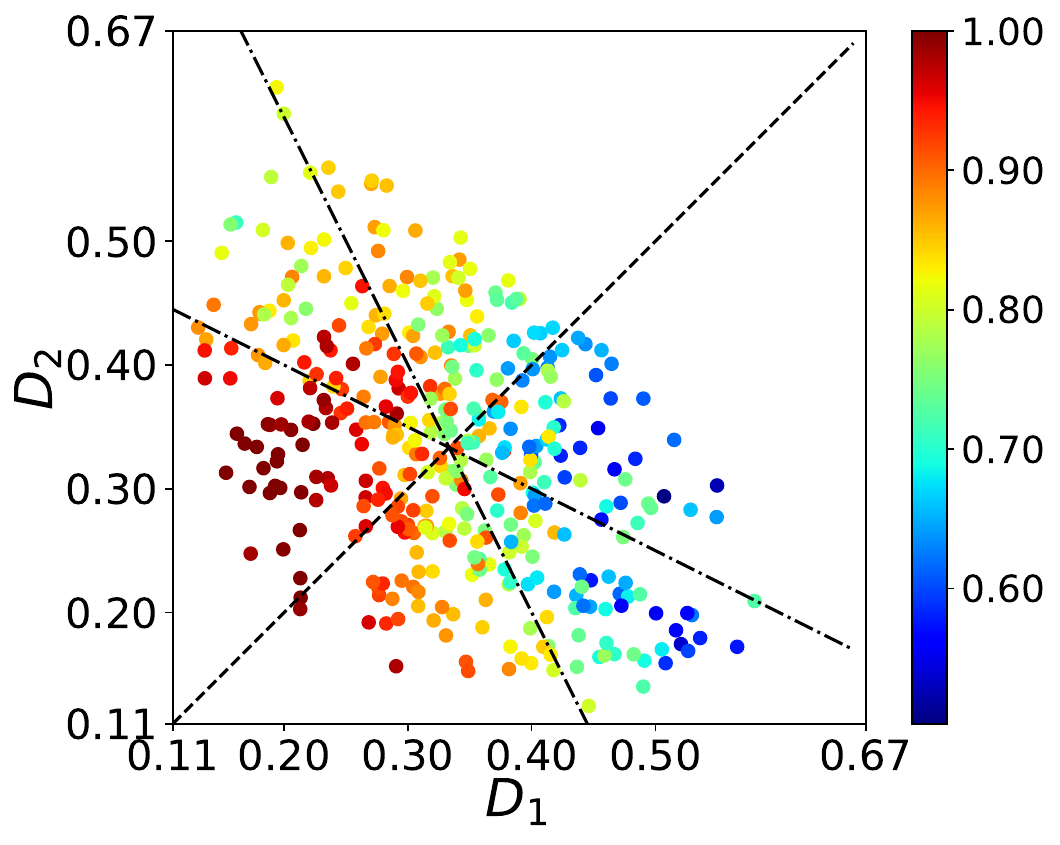}}
    \subcaptionbox{config 7}{\includegraphics[width=0.32\linewidth]{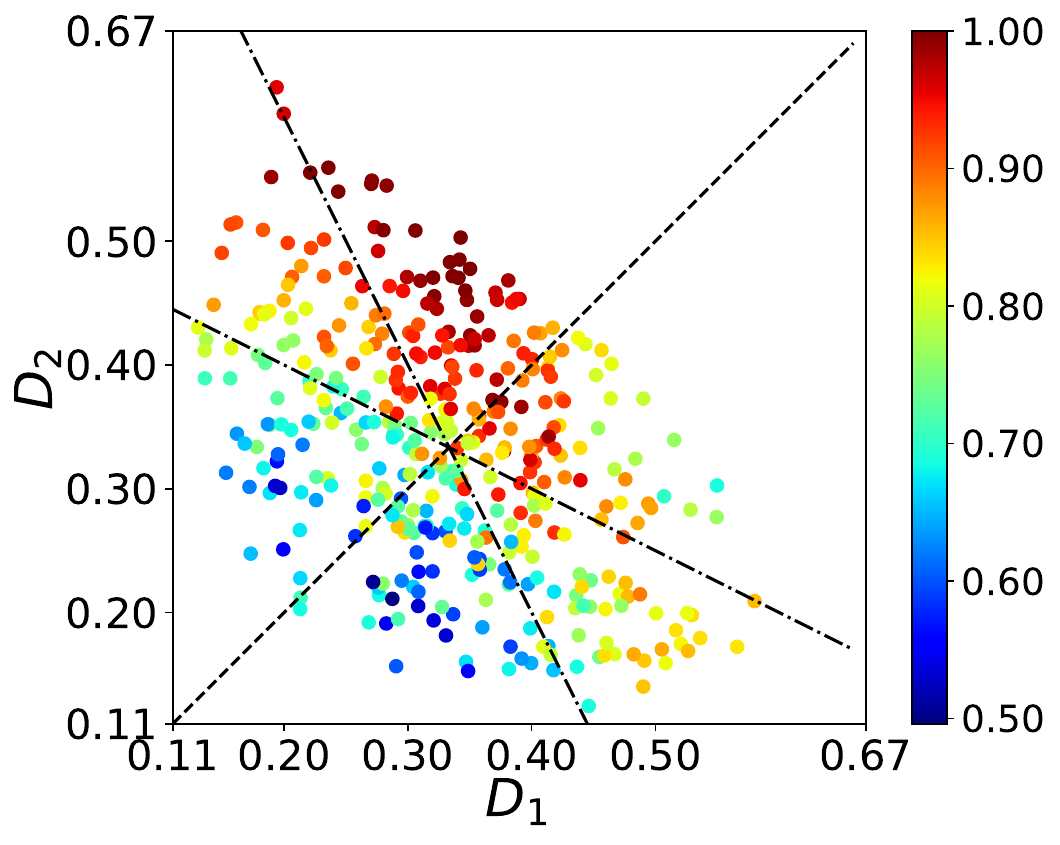}}
    \subcaptionbox{config 8}{\includegraphics[width=0.32\linewidth]{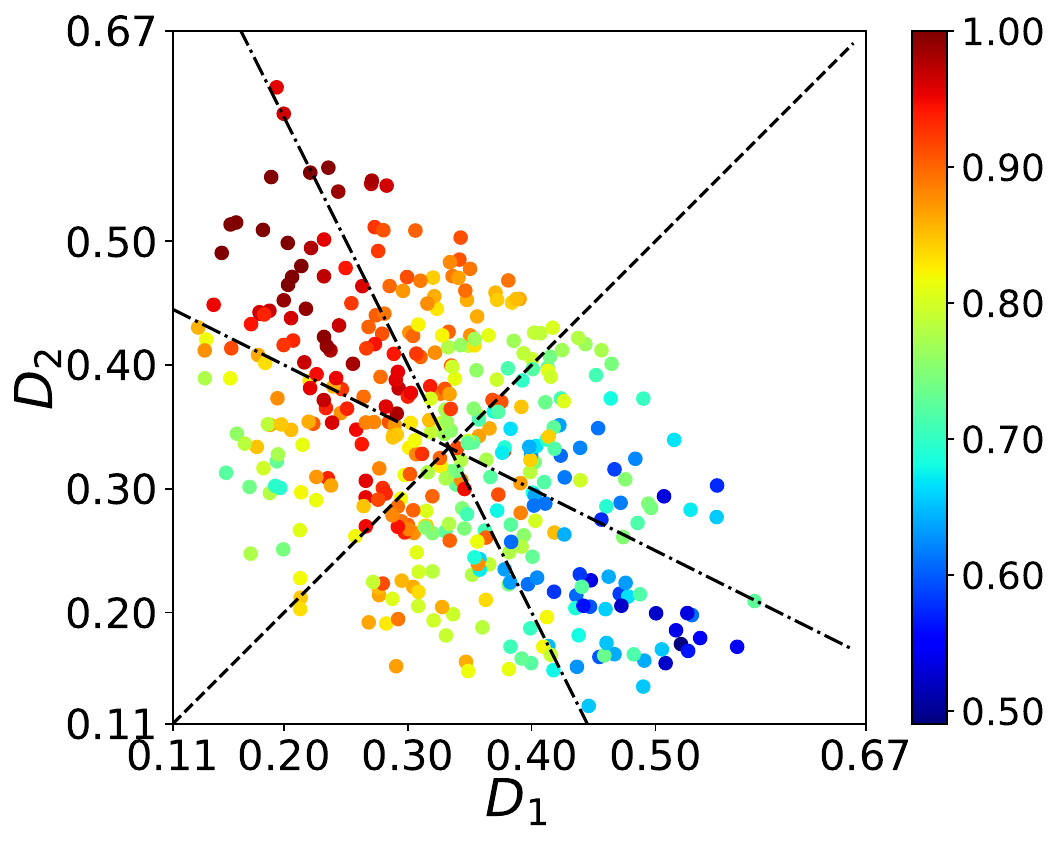}}
    \subcaptionbox{config 9}{\includegraphics[width=0.32\linewidth]{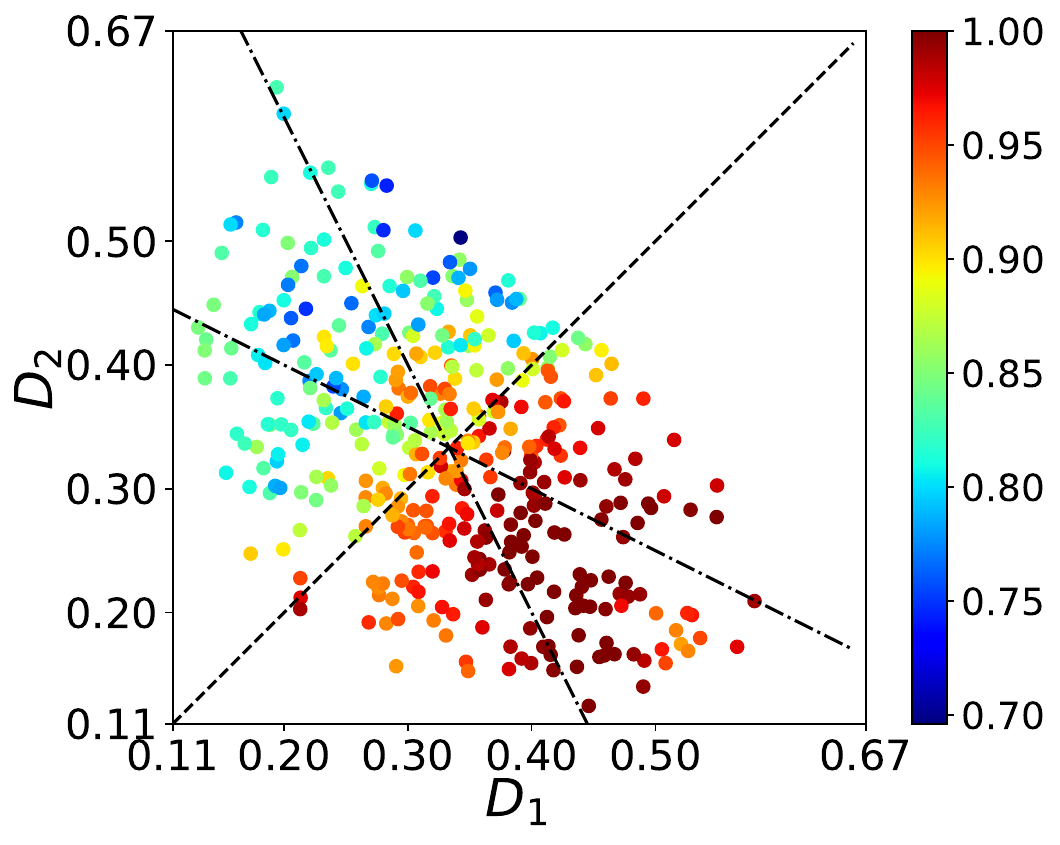}}
    \subcaptionbox{config 10}{\includegraphics[width=0.32\linewidth]{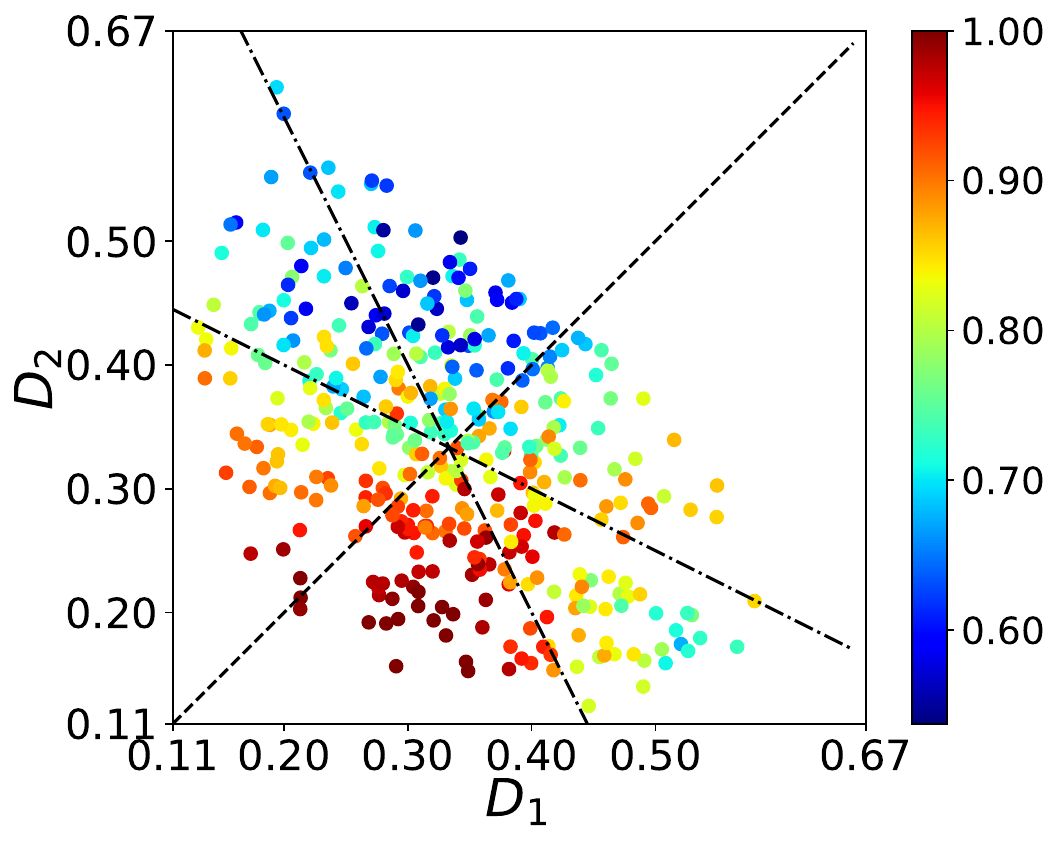}}
    \subcaptionbox{config 11}{\includegraphics[width=0.32\linewidth]{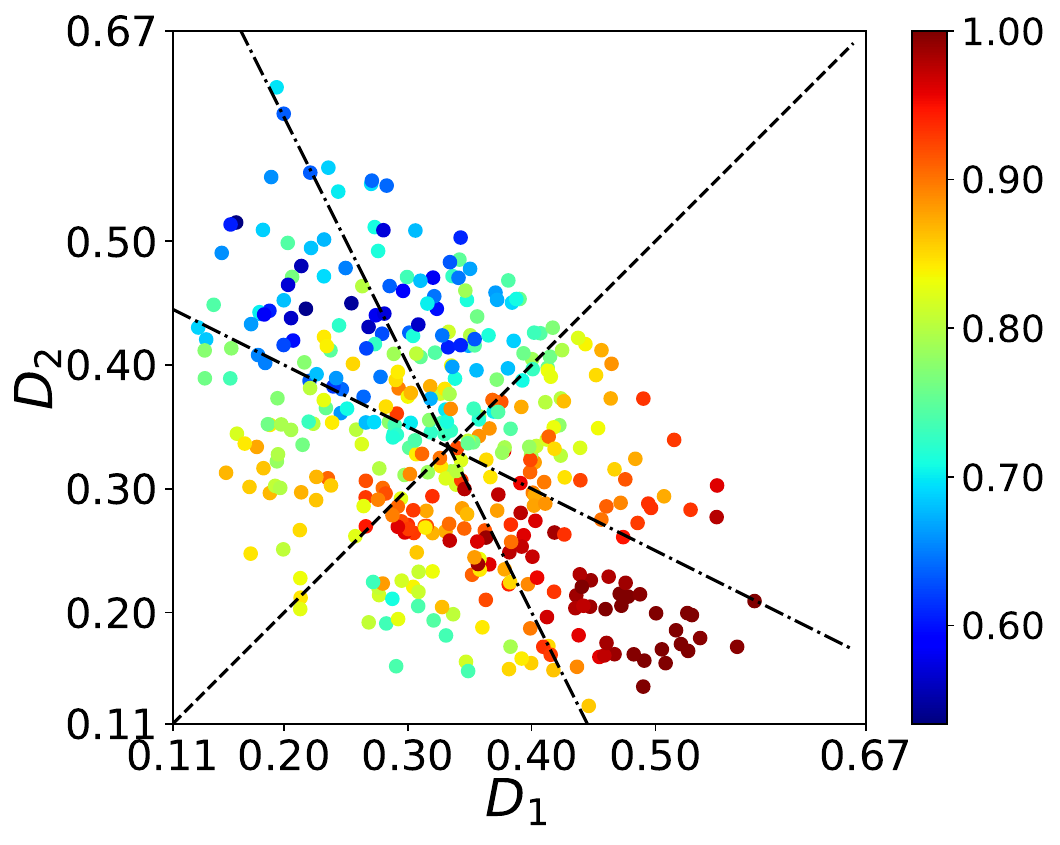}}
    \subcaptionbox{config 12}{\includegraphics[width=0.32\linewidth]{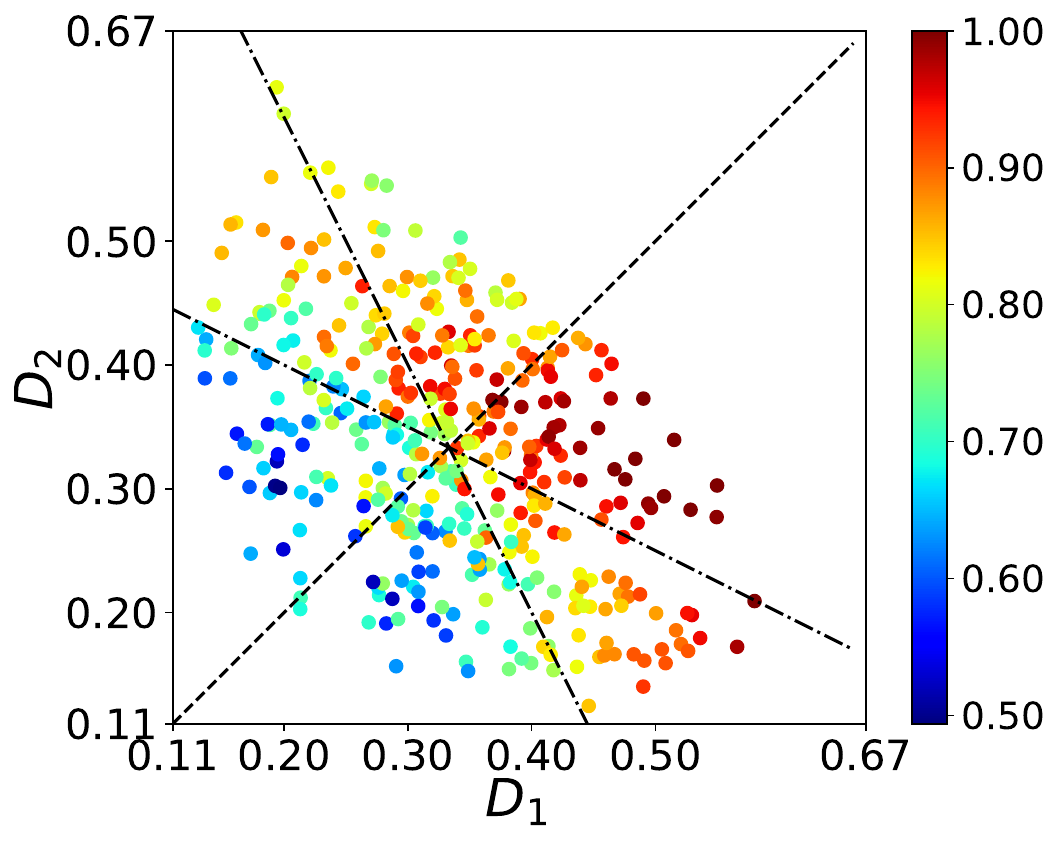}}
    \caption{Population obtained for each configuration from hypercube sampling for single split cases after feature selection defined as $D_1=d_1/\Sigma_{i=1}^3 d_i$, and $D_2=d_2/\Sigma_{i=1}^3 d_i$. The color bar shows the relative value of that configuration compared to other configuration.}
    \label{fig:scatter single3 each config}
\end{figure}

In the previous graphs, only the label for the best configuration was displayed for each disturbance. However, as mentioned before, there are 13 different configurations for each disturbance. Figure~\ref{fig:scatter single3 each config} showcases the relative importance of each configuration compared to all others across the two dimensions of features. The legend indicates the relative objective value of each configuration in comparison to all other configurations. For example, among 400 disturbance training data, where each disturbance yields 13 different configurations, if we denote the objective value of configuration $i$ for the $k$-th disturbance as $J_i^k$, the legends in the figure represent the ratio $J_i^k/\max_m J_m^k$, where $m=0,1,\dots,12$. Because of the population size, there are 400 points in each subplot.

By way of example, for instance, we can observe that configuration 0 achieves the highest value when disturbances are close to each other. In addition, as was shown in Fig~\ref{fig:singke_3_null}, classes 4 and 7 exist in the same region ($D_2>D_1>D_3$), with the only difference being whether $D_2>>D_1$ or $D_2>D_1$. The previous figure (Fig.~\ref{fig:singke_3_null}) displayed the optimal class, indicating that these classes are close to each other in the feature space ($D_1, D_2$). Now, in Fig.~\ref{fig:scatter single3 each config}, we can observe the relative importance of each configuration compared to all others across the two dimensions of disturbances.

In subplots ``e'' and ``h'', we can see that the maximum values of configuration 4 and 7 are in the top center region. Focusing on that region, it becomes apparent that as $D_2$ increases, the optimal solution generally shifts from configuration 4 to configuration 7. However, this difference is not significant when compared to all other configurations. In other words, if we are in the top center region where $D_2>D_1>D_3$, choosing either config 4 or 7 does not have a substantial impact. As a result, during the classification process, achieving high classification accuracy is not of utmost importance because obtaining an estimate of a good configuration is sufficient to yield a result with an objective value close to the optimal one.

It should be mentioned that the training data for classification is obtained by solving the OLOC problem, involving dynamics, constraints, and other factors. However, we did not include any features related to OLOC; instead, we added features based on disturbance. One of the reasons for not including more features was to avoid solving the computationally expensive OLOC problem. The goal is to obtain an estimated optimal configuration based on a given heat load. Subsequently, a designer can further investigate and include more detailed analysis to design the system.

\begin{figure}[ht!]
    \centering
    \includegraphics[width=0.8\linewidth]{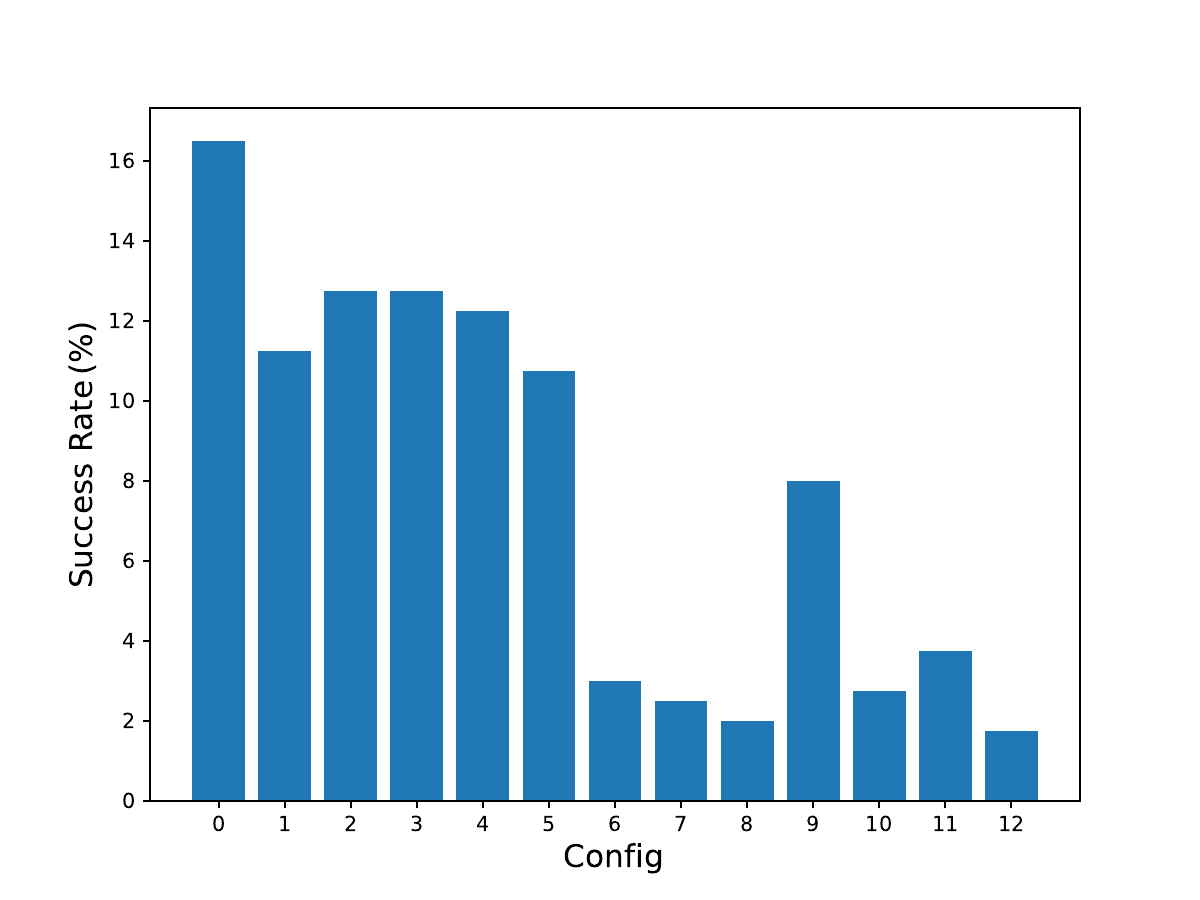}
    \caption{Single-split configuration success rate.}
    \label{fig:success rate}
\end{figure}

Among the 13 configurations, certain configurations exhibit a higher success rate (being optimal configuration) compared to others. This is illustrated in Fig.~\ref{fig:success rate} using the 400 population data. Notably, configuration 0 demonstrates the highest success rate. This aligns with part "a" of the Fig.~\ref{fig:scatter single3 each config}, where the legend shows that the difference between the minimum and maximum objective values is small (min=0.9, and max=1); in all other cases, the difference is higher. For example, in config 12, the minimum is 0.5, and the maximum is 1. This outcome is anticipated because configuration 0 comprises of three parallel branches, providing the OLOC problem with maximum flexibility in selecting the flow rate for each branch, thereby maximizing thermal endurance.

In the next step, we apply different classification methods to classify all 13 configurations based on the provided features. Table~\ref{tab:classiifcation result} presents the classification results obtained from various techniques, reporting the accuracy for both the test and training data. Among the methods tested, K-Nearest Neighbors (KNN) achieves the highest classification accuracy on both data-sets. As previously mentioned, we do not expect exceptionally high accuracy due to the nature of the problem.

\begin{figure}[ht!]
    \centering
    \subcaptionbox{MLP}{\includegraphics[width=0.49\linewidth]{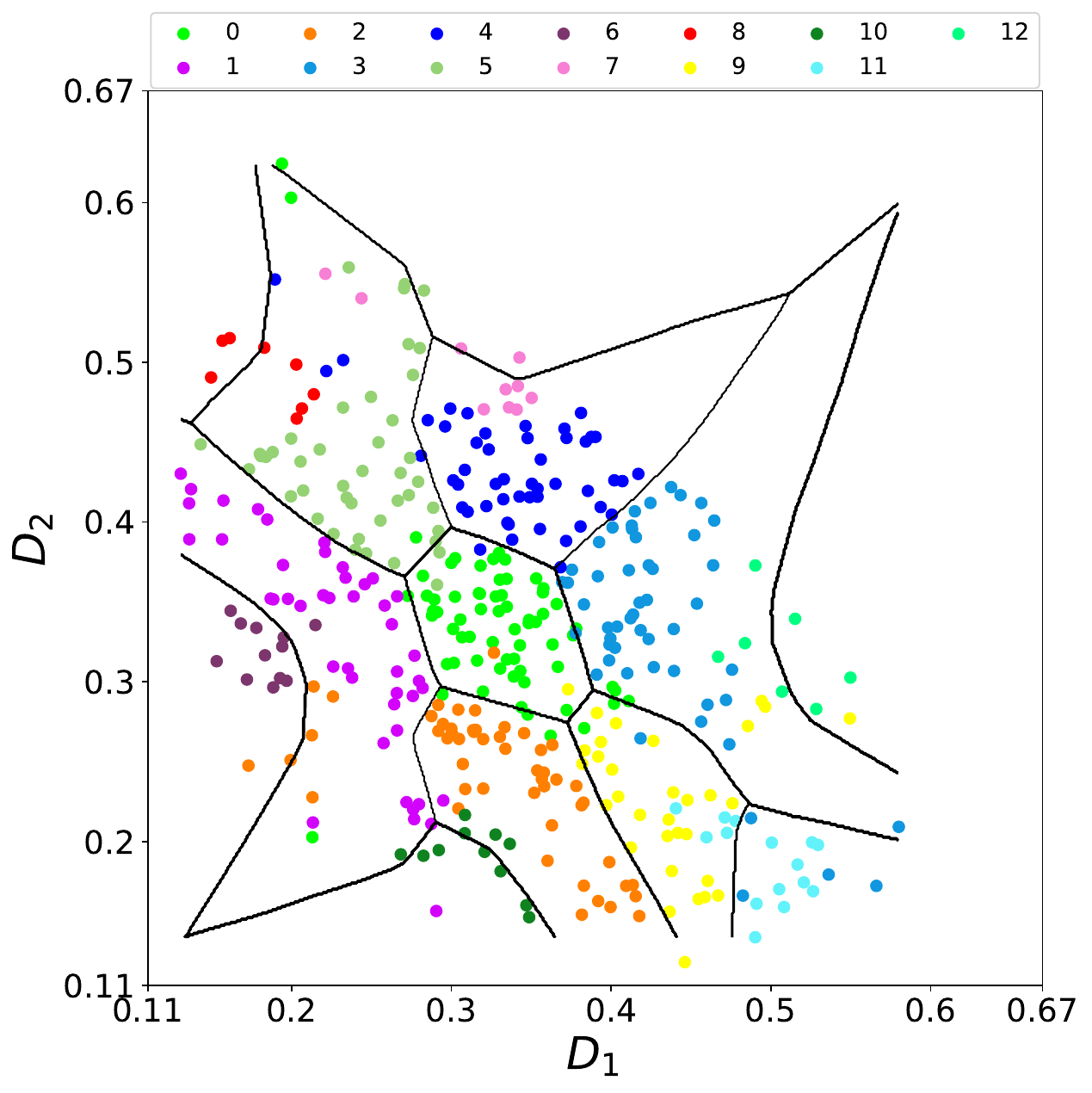}}
    \subcaptionbox{QDA}{\includegraphics[width=0.49\linewidth]{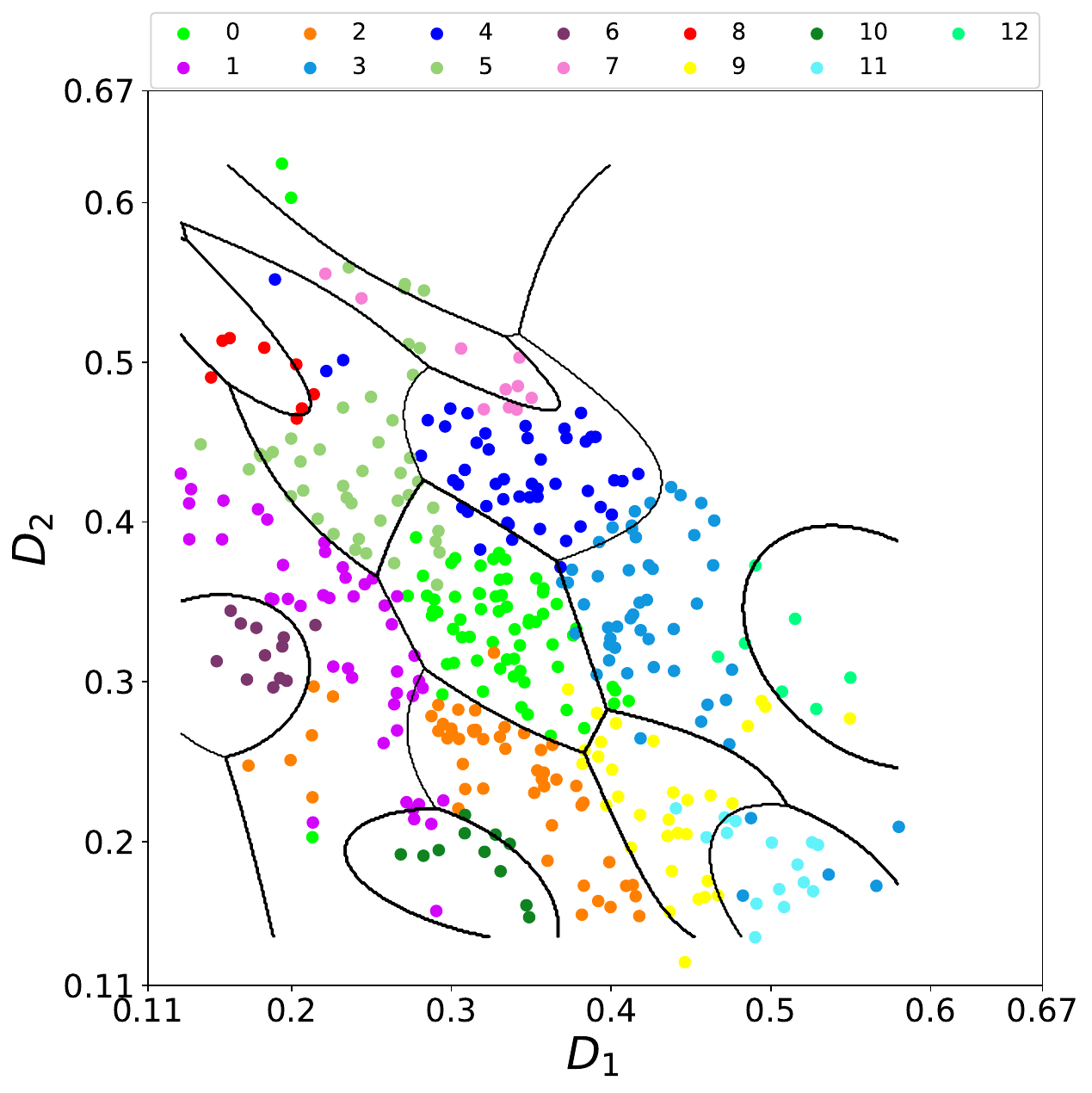}}
    \subcaptionbox{NN}{\includegraphics[width=0.49\linewidth]{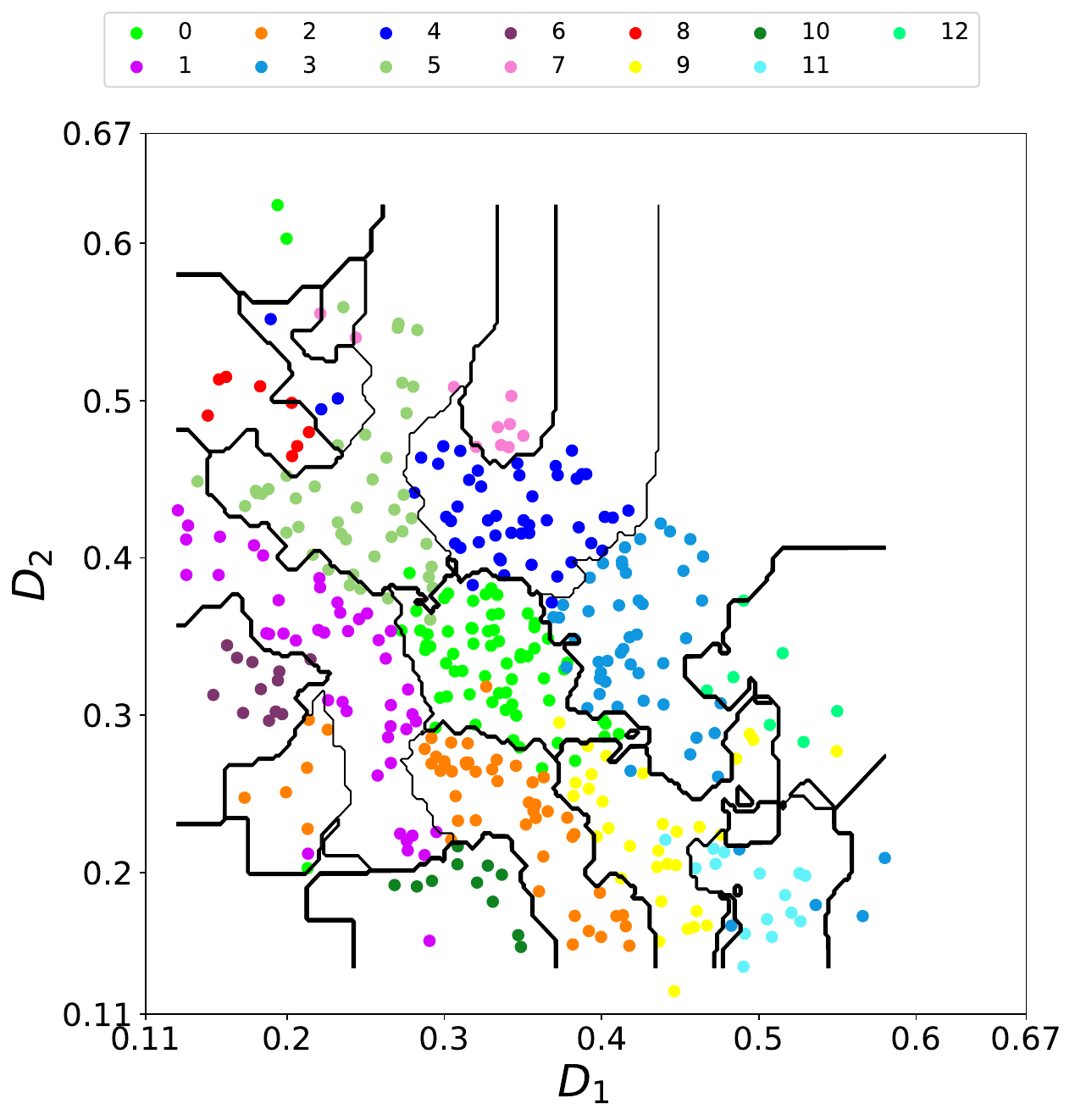}}
    \subcaptionbox{SVC linear}{\includegraphics[width=0.49\linewidth]{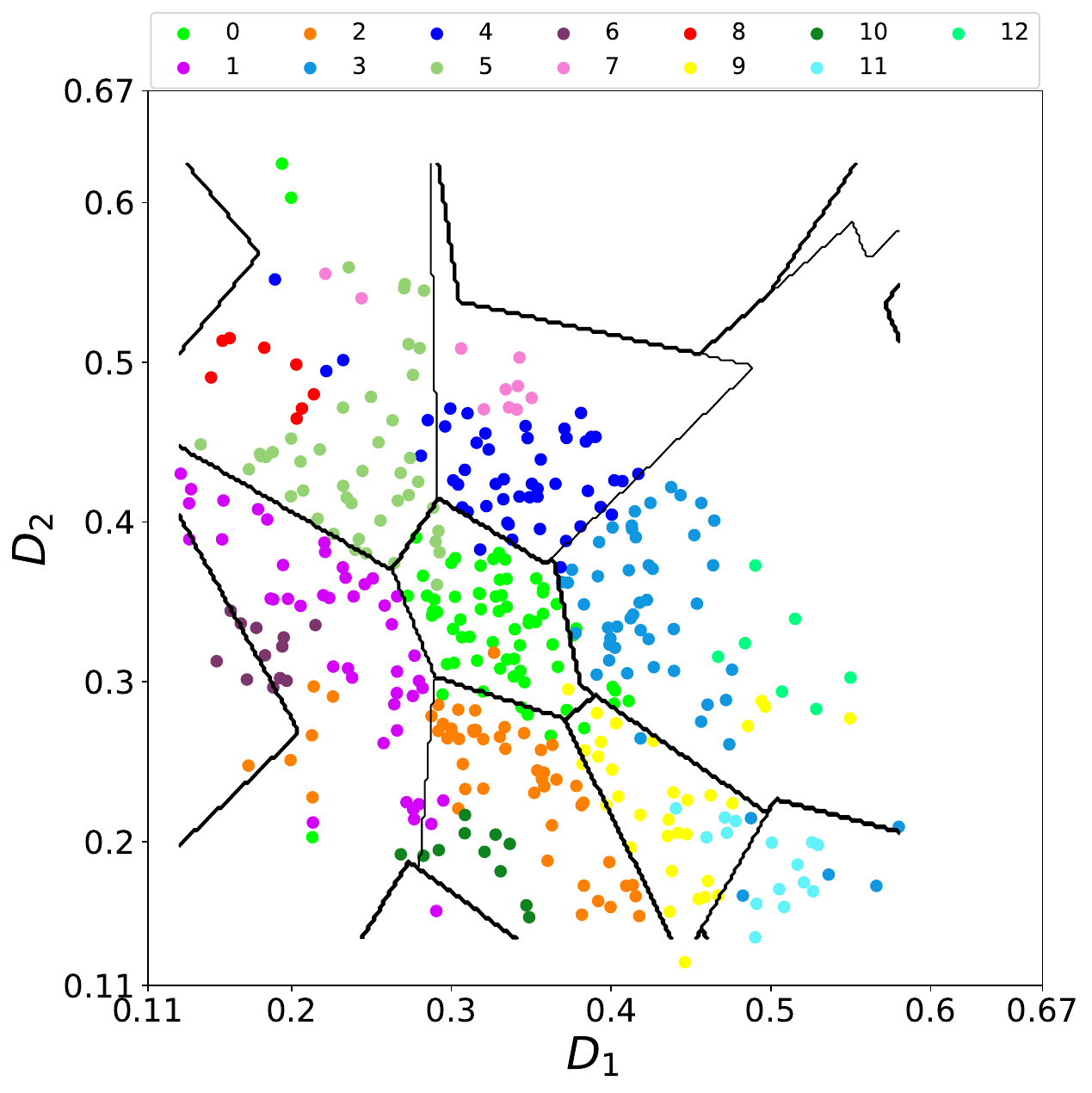}}
\caption{Compare different methods for single split cases with 3 CPHXs. The accuracy of these methods are shown in Table~\ref{tab:classiifcation result}. Here, the boundaries illustrate how each classification method assigns data to its corresponding label.}
    \label{fig: comp_methods_3_single}
\end{figure}

Figure~\ref{fig: comp_methods_3_single} depicts the boundary regions generated by four different classification methods mentioned in Table~\ref{tab:classiifcation result}. 
The boundaries here illustrate how each classification method assigns data to its respective label. The boundary changes based on the chosen algorithm, with KNN demonstrating the best results, where nodes with the same color are categorized into the same group. It's important to note, we do not claim KNN to be the absolute best method, as there are various hyper-parameters in these algorithms that can be modified. The current study serves to explore different classification methods and showcase their results. Further investigations can be pursued in future work to delve into the optimization of these methods.

\begin{table}[ht!]
\small
\centering
\caption{The classification results for different classification techniques with 3 CPHXs}
\label{tab:classiifcation result}       
\scalebox{0.7}{
\begin{tabular}{cccc}
\toprule
Method     & Acc (Test/Train)  & Method        & Acc (Test/Train)   \\ \hline
Logistic Regression        & 0.66/0.70 & Random Forest & 0.66/0.72 \\
K-Nearest Neighbours (K-NN)        & 0.78/0.93  & Naïve Bayes   & 0.70/0.75 \\
SVC (Support Vector Classifier)        & 0.70/0.76 & AdaBoost        & 0.44/0.36 \\
Kernel SVM (Support Vector Machine) & 0.72/0.82  &  Decision tree    & 0.60/0.65 \\
Quadratic Discriminant Analysis   & 0.74/0.90 & MLP           & 0.76/0.82 \\ \bottomrule
\end{tabular}
}
\end{table}

The confusion matrices for both training and test data for KNN algorithm are presented in Tables \ref{tab: rain_forest_conf_matrix_train} and \ref{tab: rain_forest_conf_matrix_test: 100 data (0.25)}, respectively. Throughout this table, each row represents the correct class (label), each column shows the prediction (predicted class), and each element indicates how often the corresponding cases have occurred. As observed, KNN performs well for both the test and training data in general.

\begin{table}[ht!]
\small
\centering
\caption{KNN confusion matrix with max depth = 3 for the training data}
\label{tab: rain_forest_conf_matrix_train}
\scalebox{0.7}{
\begin{tabular}{cccccccccccccc}
\toprule
True/Predict            & 0                          & 1                          & 2                          & 3                          & 4                          & 5                          & 6                         & 7                         & 8                         & 9                          & 10                        & 11                         & 12                        \\ \hline
\multicolumn{1}{c|}{0}  & \cellcolor[HTML]{C0C0C0}51 & 0                          & \cellcolor[HTML]{96FFFB}1  & 0  & 0  & \cellcolor[HTML]{96FFFB}1                          & 0                         & 0                         & 0                         & 0                          & 0                         & 0                          & 0                         \\
\multicolumn{1}{c|}{1}  & 0                          & \cellcolor[HTML]{C0C0C0}38 & 0  & 0                          & 0                          & 0                          & 0                         & 0                         & 0                         & 0                          & \cellcolor[HTML]{96FFFB}1                          & 0                          & 0                         \\
\multicolumn{1}{c|}{2}  & \cellcolor[HTML]{96FFFB}1  & \cellcolor[HTML]{96FFFB}1  & \cellcolor[HTML]{C0C0C0}44 & 0                          & 0                          & 0                          & 0                         & 0                         & 0                         & 0                          & 0                         & 0                          & 0                         \\
\multicolumn{1}{c|}{3}  & \cellcolor[HTML]{96FFFB}2                          & 0                          & 0                          & \cellcolor[HTML]{C0C0C0}39 & 0                          & 0                          & 0                         & 0                         & 0                         & \cellcolor[HTML]{96FFFB}1  & 0                         & \cellcolor[HTML]{96FFFB}2  & 0                         \\
\multicolumn{1}{c|}{4}  & \cellcolor[HTML]{96FFFB}1  & 0                          & 0                          & \cellcolor[HTML]{96FFFB}1                          & \cellcolor[HTML]{C0C0C0}44 & \cellcolor[HTML]{96FFFB}1  & 0                         & 0                         & 0                         & 0                          & 0                         & 0                          & 0                         \\
\multicolumn{1}{c|}{5}  & \cellcolor[HTML]{96FFFB}1  & \cellcolor[HTML]{96FFFB}1                         & 0                          & 0                          & 0  & \cellcolor[HTML]{C0C0C0}30 & 0                         & \cellcolor[HTML]{96FFFB}1                         & 0                         & 0                          & 0                         & 0                          & 0                         \\
\multicolumn{1}{c|}{6}  & 0                          & \cellcolor[HTML]{96FFFB}1 & 0                          & 0                          & 0                          & 0                          & \cellcolor[HTML]{C0C0C0}11 & 0                         & 0                         & 0                          & 0                         & 0                          & 0                         \\
\multicolumn{1}{c|}{7}  & 0                          & 0                          & 0                          & 0                          & \cellcolor[HTML]{96FFFB}2  & \cellcolor[HTML]{96FFFB}1                          & 0                         & \cellcolor[HTML]{C0C0C0}6 & 0                         & 0                          & 0                         & 0                          & 0                         \\
\multicolumn{1}{c|}{8}  & 0                          & 0                          & 0                          & 0                          & 0                          & 0  & 0                         & 0                         & \cellcolor[HTML]{C0C0C0}5 & 0                          & 0                         & 0                          & 0                         \\
\multicolumn{1}{c|}{9}  & 0                          & 0                          & \cellcolor[HTML]{96FFFB}1       & 0  & 0                          & 0                          & 0                         & 0                         & 0                         & \cellcolor[HTML]{C0C0C0}27 & 0                         & \cellcolor[HTML]{96FFFB}1                          & \cellcolor[HTML]{96FFFB}1                         \\
\multicolumn{1}{c|}{10} & 0                          & 0                          & 0 & 0                          & 0                          & 0                          & 0                         & 0                         & 0                         & 0                          & \cellcolor[HTML]{C0C0C0}10 & 0                          & 0                         \\
\multicolumn{1}{c|}{11} & 0                          & 0                          & 0                          & 0                          & 0                          & 0                          & 0                         & 0                         & 0                         & \cellcolor[HTML]{96FFFB}1  & 0                         & \cellcolor[HTML]{C0C0C0}14 & 0                         \\
\multicolumn{1}{c|}{12} & 0                          & 0                          & 0                          & \cellcolor[HTML]{96FFFB}1  & 0                          & 0                          & 0                         & 0                         & 0                         & \cellcolor[HTML]{96FFFB}1                          & 0                         & 0                          & \cellcolor[HTML]{C0C0C0}5 \\ \bottomrule
\end{tabular}
}
\end{table}

\begin{table}[ht!]
\small
\centering
\caption{KNN confusion matrix with max depth = 3 for the test data}
\label{tab: rain_forest_conf_matrix_test: 100 data (0.25)}
\scalebox{0.9}{
\begin{tabular}{ccccccccccc}
\toprule
True/Predict            & 0                          & 1                          & 2                         & 3                         & 4                          & 5                          & 7                         & 8                         & 9 &10           \\ \hline
\multicolumn{1}{c|}{0}  & \cellcolor[HTML]{C0C0C0}11 & 0                          & \cellcolor[HTML]{96FFFB}1 & 0 & 0                          & 0                          & 0                         & 0                         & \cellcolor[HTML]{96FFFB}1                         & 0     \\
\multicolumn{1}{c|}{1}  & 0                          & \cellcolor[HTML]{C0C0C0}5 & \cellcolor[HTML]{96FFFB}1                         & 0                         & 0                          & 0                          & 0                         & 0                         & 0                         & 0                       \\
\multicolumn{1}{c|}{2}  & 0                          & 0  & \cellcolor[HTML]{C0C0C0}5 & 0                         & 0                          & 0                          & 0                         & 0                         & 0                         & 0   \\
\multicolumn{1}{c|}{3}  & 0  & 0                          & 0                         & \cellcolor[HTML]{C0C0C0}7 & 0                          & 0                          & 0                         & 0                         & 0                         & 0  \\
\multicolumn{1}{c|}{4}  & 0                          & 0                          & 0                         & 0                         & \cellcolor[HTML]{C0C0C0}2 & 0  & 0                         & 0                         & 0                         & 0   \\
\multicolumn{1}{c|}{5}  & 0                          &      \cellcolor[HTML]{96FFFB}2                     & 0                         & 0                         & \cellcolor[HTML]{96FFFB}1                          & \cellcolor[HTML]{C0C0C0}8 & 0                         & 0                         & 0                         & 0   \\
\multicolumn{1}{c|}{7}  & 0                          & 0                          & 0                         & 0                         & 0                          & \cellcolor[HTML]{96FFFB}1                          & \cellcolor[HTML]{C0C0C0}0 & 0                         & 0                         & 0     \\
\multicolumn{1}{c|}{8}  & 0                          & 0                          & 0                         & 0                         & 0  & 0                          & 0                         & \cellcolor[HTML]{C0C0C0}3 & 0                         & 0  \\
\multicolumn{1}{c|}{9}  & \cellcolor[HTML]{96FFFB}1                          & 0                          & 0                         & 0                         & 0                          & 0  & 0                  &0       & \cellcolor[HTML]{C0C0C0}1                        & 0          \\
\multicolumn{1}{c|}{10}  & 0 & 0                          & 0                         & 0                         & 0                          & 0                          & 0                         & 0                         & 0                         & \cellcolor[HTML]{C0C0C0}1     \\
\bottomrule
\end{tabular}
}
\end{table}

\begin{figure}[ht!]
    \centering
    \includegraphics[width=1.0\linewidth]{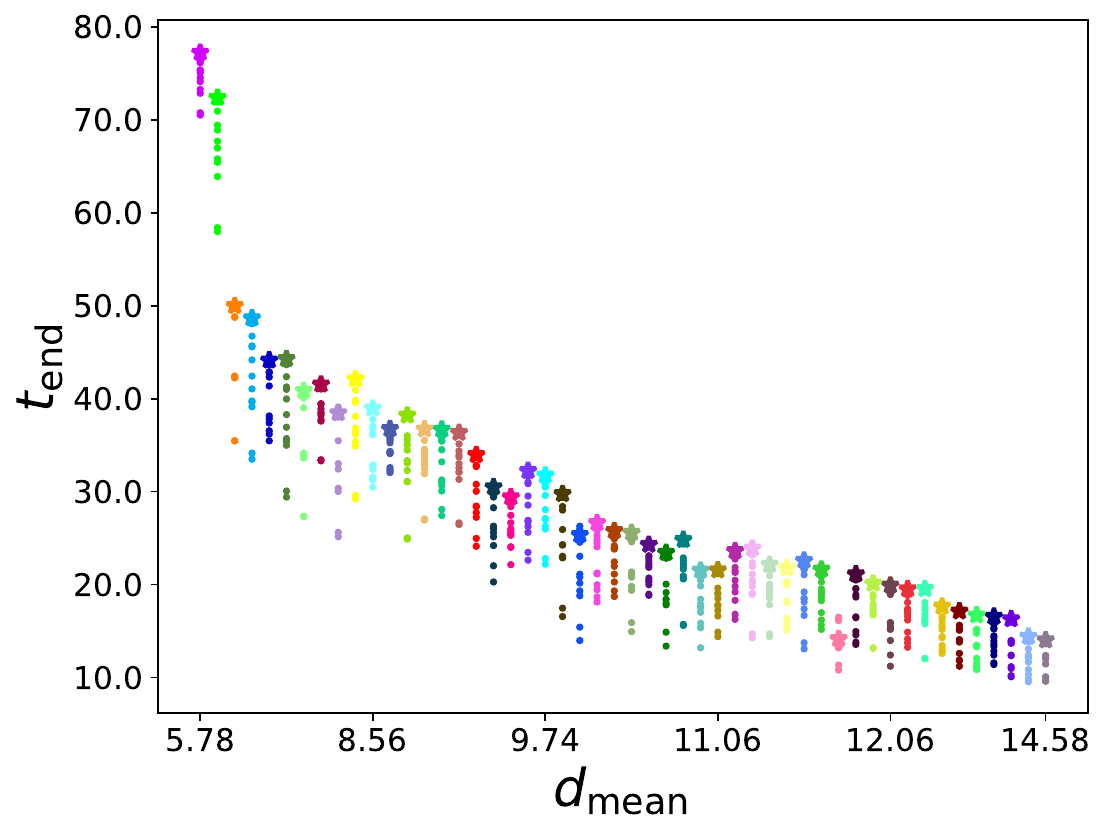}
    \caption{Objective value comparison of estimated optimal configuration ($\ast$) with Other populations ($\cdot$) for each disturbance on Test data for single split cases with 3 nodes.}
    \label{fig:3_nodes_KNN}
\end{figure}

Figure~\ref{fig:3_nodes_KNN} compares the objective values of the predicted configurations using KNN with the objective values of all other configurations. The test data consists of a population size of 50, resulting in 50 different disturbances, with each disturbance having 13 different configurations. For each disturbance, we predict the label of the configuration with the best thermal endurance, i.e. maximum objective value. The objective values associated with these predicted labels are denoted by asterisks (*) in the figure. The results show that even if the predicted labels do not match the true labels, their objective are in close proximity to the true labels and can yield similar objective values. In this figure, each color represents a unique disturbance defined by $[d_{1},d_{2}, d_{3}]$. The x-axis represents the average value of these disturbances ($d_{\mathrm{mean}}=(d_1+d_2+d_3)/3$), while the y-axis represents the objective function value ($t_{\mathrm{end}}$). Consequently, for each mean disturbance, there are 13 different points on the y-axis. For many cases, the predicted label (the case marked with an asterisk) is located at the top among all 13 cases and has an objective value close to the optimal value. This indicates that, despite the KNN accuracy not being particularly high (0.78 in this case on test data), the predicted label is quite close to the optimal label. Therefore, the data obtained through this classification process can provide engineers with an estimation of configurations that may yield optimal solutions. An additional intuition that can be obtained from this figure is that, as demonstrated, generally an increase in the mean disturbance, as we move along the higher values of the x-axis, results in a decrease in the objective value.

\subsection{Multi-Split Graphs with 3 Nodes }
\label{sec: CPHx with 3 nodes- Multi Split}

\begin{figure}[ht!]
    \centering
    \includegraphics[width=0.8\linewidth]{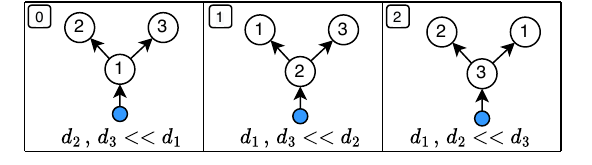}
    \caption{Unique multi-split structures with 3 heat generating components. Also presented is the heat load condition for which each configuration is the optimal structure.}
    \label{fig:multi_3_configs}
\end{figure}

\begin{figure}[ht!]
    \centering
    \includegraphics[width=1.0\linewidth]{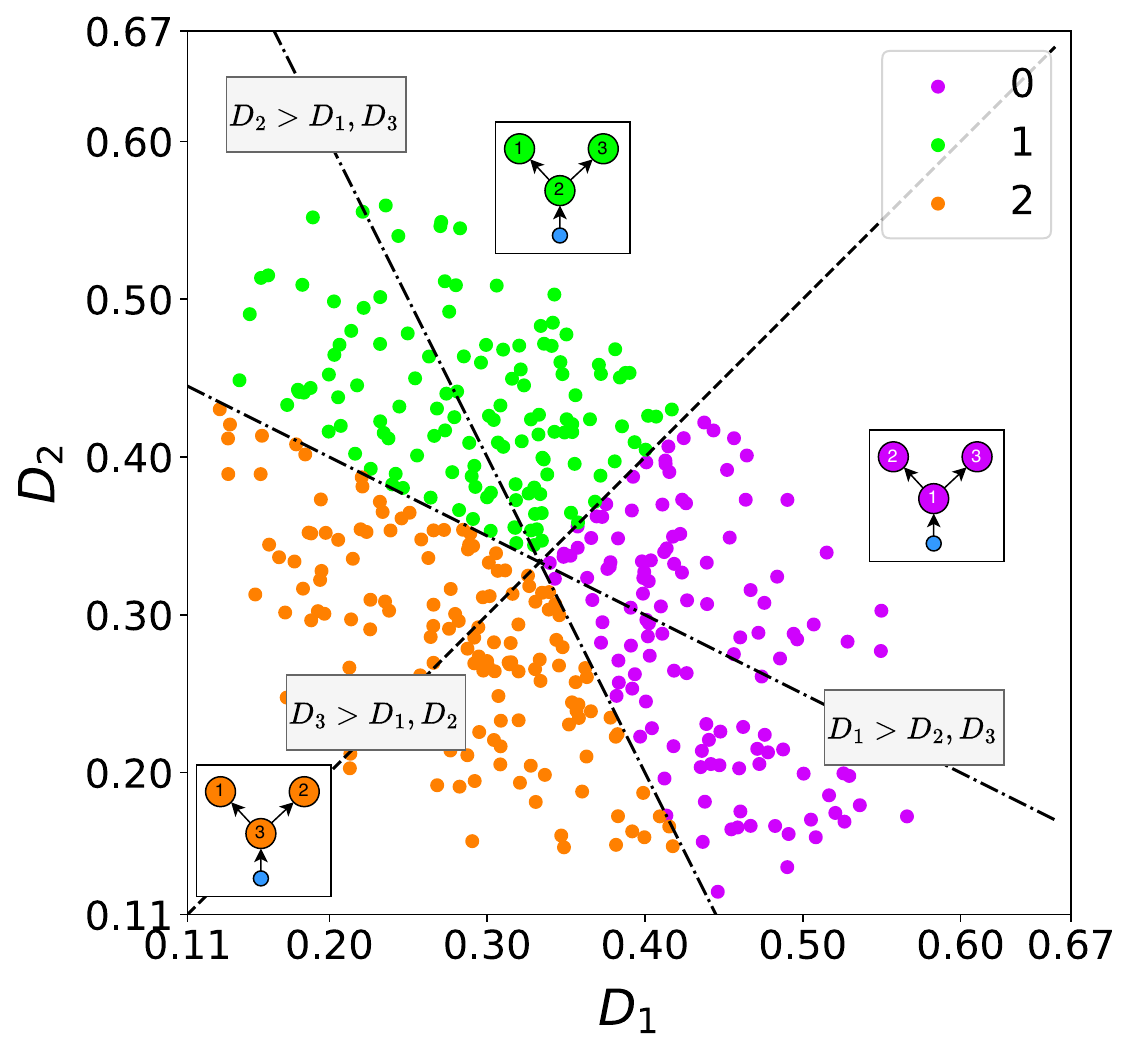}
    \caption{Population obtained from hypercube sampling after feature selection defined as $D_1=d_1/\Sigma_{i=1}^3 d_i$, and $D_2=d_2/\Sigma_{i=1}^3 d_i$, for multi-split cases.}
    \label{fig:scatter multi3 Pop}
\end{figure}

\begin{figure}[ht!]
    \centering
    \subcaptionbox{config 0}{\includegraphics[width=0.49\linewidth]{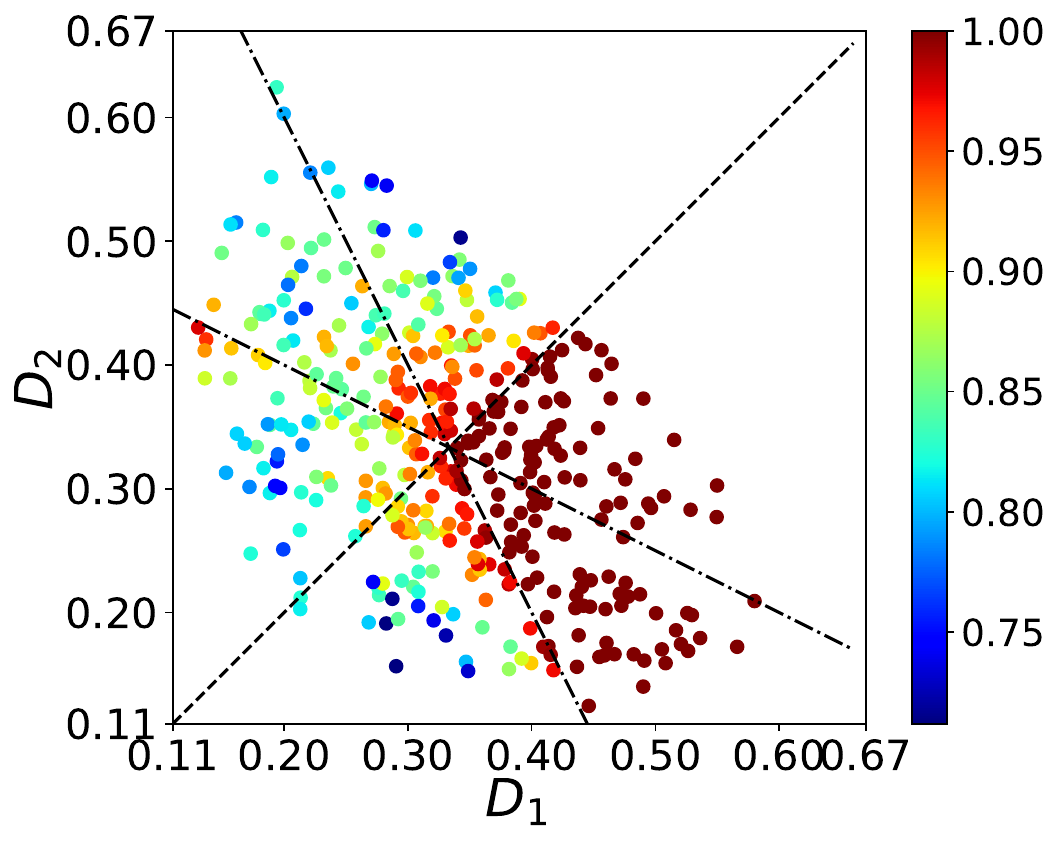}}
    \subcaptionbox{config 1}{\includegraphics[width=0.49\linewidth]{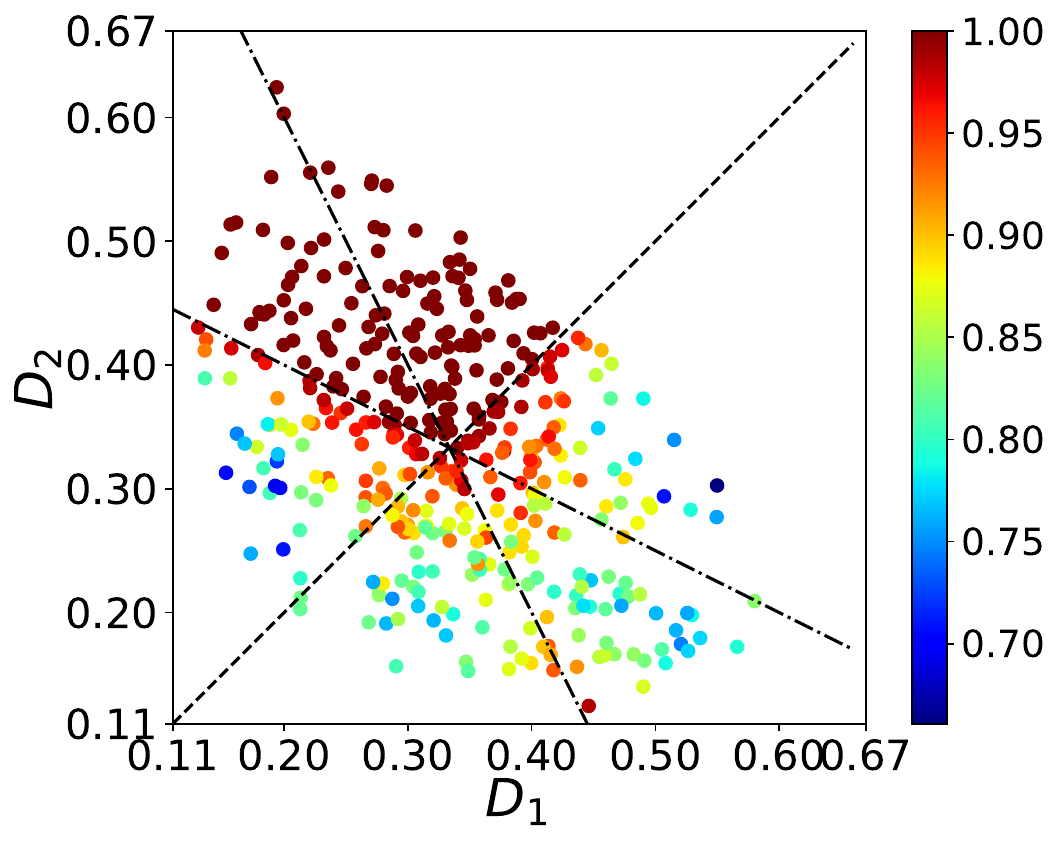}}
    \subcaptionbox{config 2}{\includegraphics[width=0.49\linewidth]{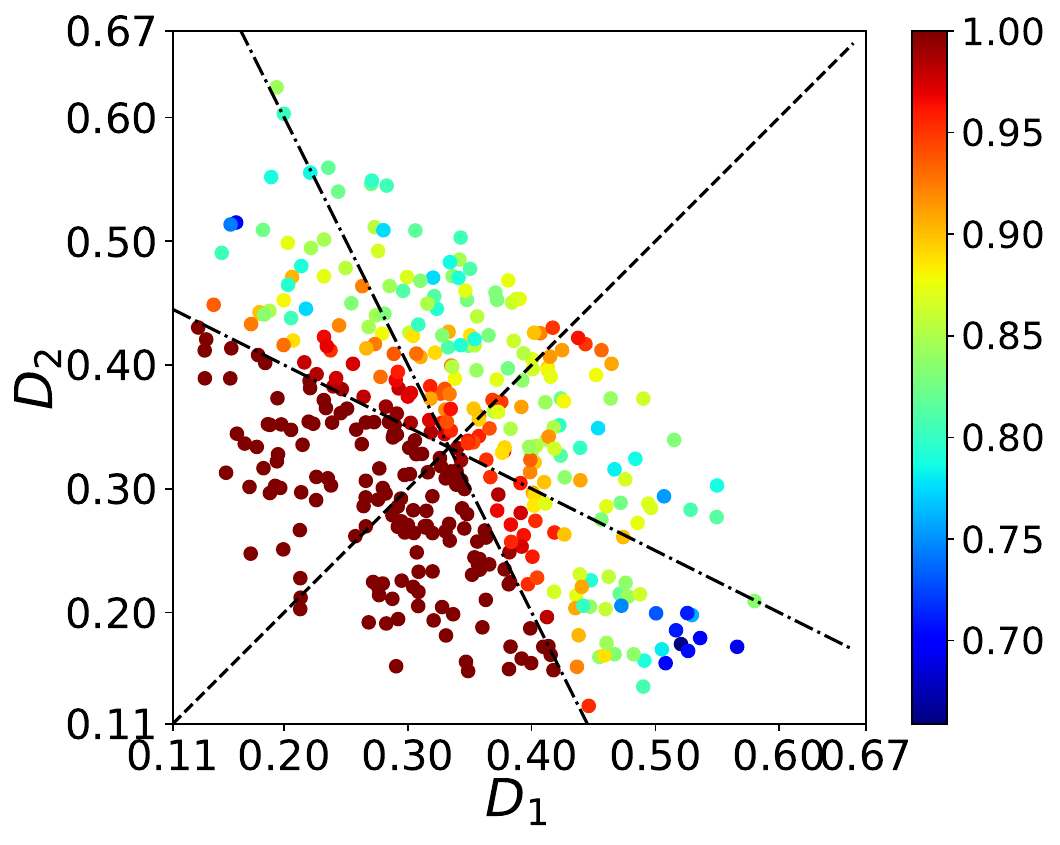}}
    \caption{Population obtained from hypercube sampling after feature selection defined as $D_1=d_1/\Sigma_{i=1}^3 d_i$, and $D_2=d_2/\Sigma_{i=1}^3 d_i$, for multi-split cases. The color bar shows the relative performance value of that configuration comparing to other configurations.}
    \label{fig: scatter multi3 configs}
\end{figure}

We now consider multi-split configurations with 3 nodes. Figure~\ref{fig:multi_3_configs} shows these three cases along with the condition in which that configuration performs the best. The parameters for training are shown in Table~\ref{tab:alg_params_3odes_multi}. Similar to the previous section, two features ($D_1$ and $D_2$) are taken into account. Figure~\ref{fig:scatter multi3 Pop} shows the best candidate from all populations and their corresponding configuration label. According to the 2D plots, these cases are clearly separable. For example, for configuration 0 we have  $d_1>>d_2,d_3$. Also, the first 2D plot shows that configuration 0 is completely distinguishable from the other two classes when $d_{12}>d_{1/3},d_{2,3}$. Figure~\ref{fig: scatter multi3 configs} also shows what happens if we move in any direction for each candidate.

\begin{table}[ht!]
\small
\centering
\caption{Values of parameters in  Algorithm~\ref{Training Procedure} that are used in Sec.\ref{sec: CPHx with 3 nodes- Multi Split} for multiple split cases with 3 CPHXS}
\label{tab:alg_params_3odes_multi}       
\scalebox{1.0}{
\begin{tabular}{cccc}
\toprule
param     & value  & param        & value  \\ \hline
$n_{\mathrm{pop}}$   & 400 & $d_{\mathrm{range}}$ & $[4,16]$kW \\
$n_{\mathrm{nodes}}$ & 3  & ${n_{\mathrm{conf}}}$   & 3 \\
$n_{\mathrm{f}}$ & 3 & $D$       & $d_i/\Sigma d_i$ \\
$n_{\mathrm{train}}$ & 350 & $n_{\mathrm{test}}$ &  50 \\
Sampling method   & LHS & & \\
\bottomrule
\end{tabular}
}
\end{table}

\begin{figure}[ht!]
    \centering
    \includegraphics[width=1.0\linewidth]{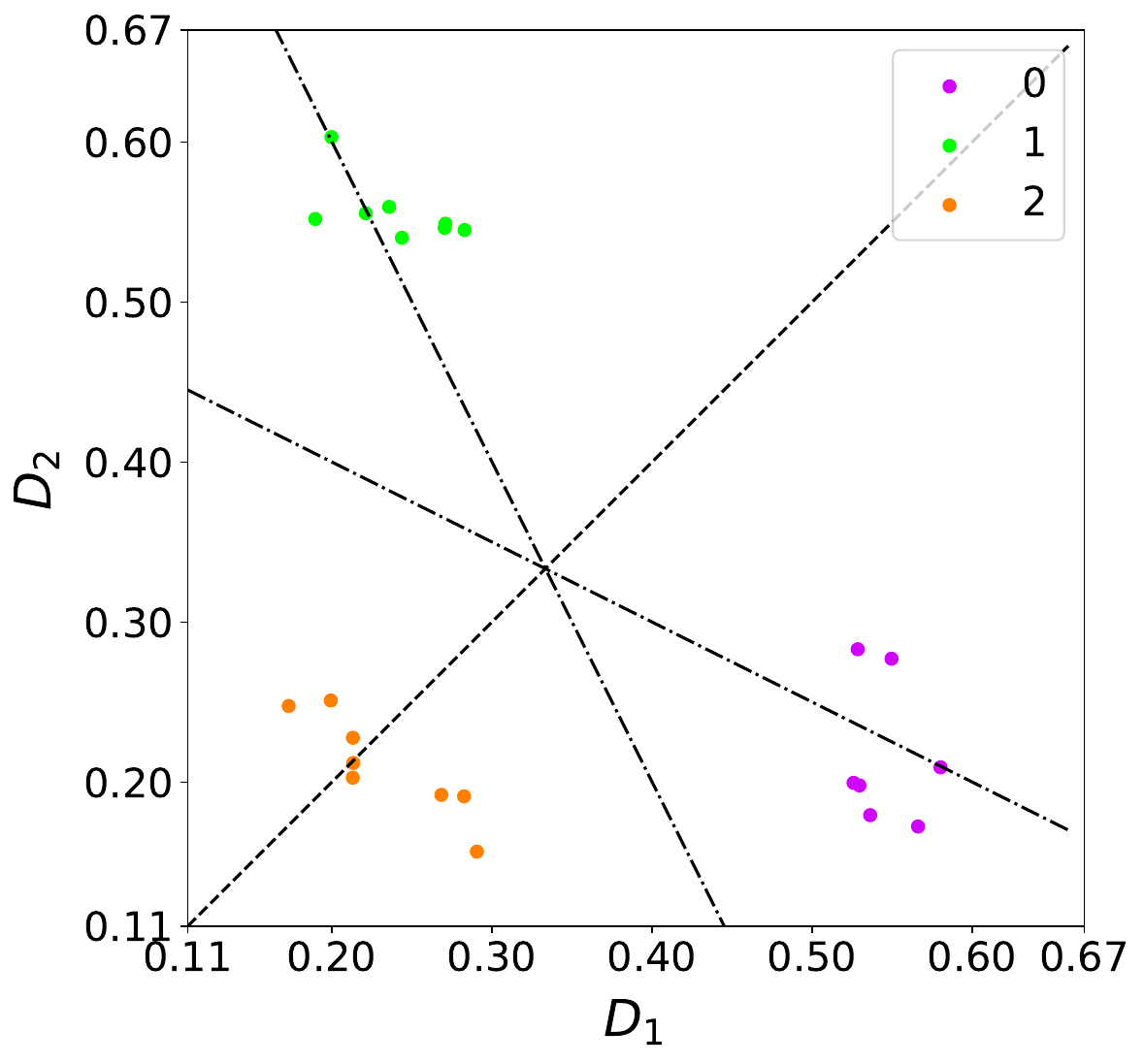}
    \caption{The region in feature space at which multi-split cases shown in Fig.~\ref{fig:multi_3_configs} have better performances than single-split cases shown in Fig.~\ref{fig:Single_3_configs}.}
    \label{scatter multi3 better than single_1}
\end{figure}

Figure~\ref{scatter multi3 better than single_1} shows instances where the multi-split configurations are more effective than the single-split configurations. These cases correspond to extreme disturbances, for example cases with $d_1>>d_2,d_3$. One reason behind this is that in such cases, the node with the maximum heat load is connected directly to the pump, receiving the maximum flow rate, which is equal to the pump flow rate. This is achieved in both multi-split configurations and in the single-split case where all nodes are in series. However, in the multi-split configuration, the controller has the freedom to adjust the flow rate in the subsequent two branches. In contrast, in the single-series case, the controller lacks authority, resulting in identical flow rates across all nodes, which are equal to the pump flow rate. This is one of the reasons why the multi-split case outperforms the single-split case.

\subsection{Single-Split Graphs with 4 Nodes}
\label{sec: CPHx with 4 nodes- Single}

In this section, we focus on the classification of single-split cases with 4 nodes. A total of 73 different configurations are considered here. The population of disturbances is generated in such a manner that the following condition holds: $d_1 \ge d_2 \ge d_3 \ge d_4$. 
Building upon the classification methods used in previous sections, we classify these configurations based on the normalized disturbances ($D_i=d_i/\Sigma d_i$ for $i \in \{1,2,3\}$). However, in this section, we introduced an additional feature: $D_4=d_1/10kW$. We recognized the importance of incorporating the magnitude of the first node (experiencing the largest disturbance) to achieve reasonable accuracy. Therefore, alongside the normalized features used in the previous section, we included the feature $D_4$ in our analysis.

\begin{figure}[ht!]
    \centering
    \includegraphics[width=1.0\linewidth]{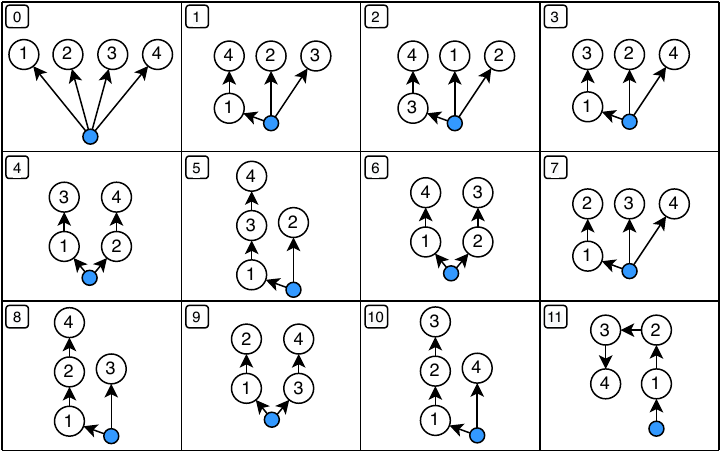}
    \caption{
    Eleven unique configurations out of 73 that identified as optimal based on the disturbance population size of 200. The disturbance population is obtained by satisfying the following constraint: $d_1 > d_2 > d_3 > d_4$}
    \label{fig:Single_4_configs}
\end{figure}

The training parameters are presented in Table~\ref{tab:alg_params_4odes_single}. Upon analyzing the true labels from the complete dataset, we discovered that only 12 distinct categories (out of 73) yielded the best outcome. In other words, upon examining all 73 configurations across the entire training population, it was observed that only 12 unique configurations yielded the maximum objective value, while all other configurations had inferior objective values. This implies that out of the 200 diverse populations, each consisting of 73 different cases, only 12 configurations (illustrated in Fig.~\ref{fig:Single_4_configs}) were labeled as optimal solutions. Notably, in all of these cases, node 1, which possesses the highest heat load, is situated nearest to the pump and receives a relatively cooler fluid flow. These 12 classes were derived under the assumption we maintained during training, where the distances were ordered as $d_1 \ge d_2 \ge d_3 \ge d_4$. Modifying this assumption would generate different optimal configurations, but the underlying concept would remain the same. In other words, the node with the maximum heat load would still be positioned closest to the tank. Therefore, the assumption we employed to generate sample points does not limit the range of configurations. We made this assumption specifically to reduce the number of labels (different graph configurations) from 73 to 12, thereby facilitating a more convenient analysis of the results.

\begin{table}[ht!]
\small
\centering
\caption{Values of parameters in  Algorithm~\ref{Training Procedure} that are used in Sec.~\ref{sec: CPHx with 4 nodes- Single} for single-split cases with 4 CPHXs}
\label{tab:alg_params_4odes_single}       
\scalebox{0.87}{
\begin{tabular}{cccc}
\toprule
param     & value  & param        & value  \\ \hline
$n_{\mathrm{pop}}$   & 200 & $d_{\mathrm{range}}$ & $[4,16]$kW \\
$n_{\mathrm{nodes}}$ & 4  & ${n_{\mathrm{conf}}}$   & 73 \\
$n_{\mathrm{f}}$ & 4 & $D$       & $[d_i/\Sigma d_i, d_0/10kW]$ \\
$n_{\mathrm{train}}$ & 175 & $n_{\mathrm{test}}$ &  25 \\
Sampling method   & Random & Condition & $d_1 \ge d_2 \ge d_3 \ge d_4$\\
\bottomrule
\end{tabular}
}
\end{table}

\begin{figure}[ht!]
    \centering
    \includegraphics[width=1.0\linewidth]{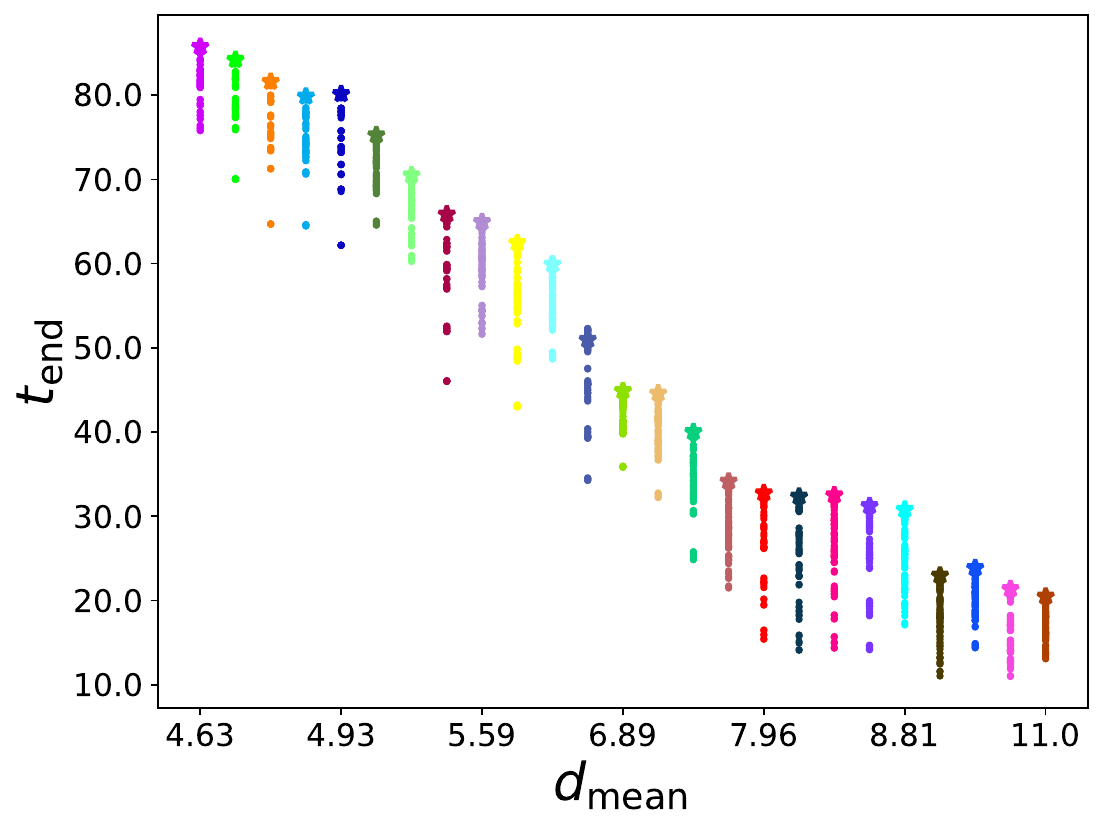}
    \caption{Objective value comparison of estimated optimal configuration ($\ast$) with Other populations ($\cdot$) for each disturbance on Test data for single split cases with 4 nodes.}
    \label{fig:4_nodes_KNN}
\end{figure}

Table ~\ref{tab:classiifcation result-4nodes} presents the accuracy of various classification methods on both the training and test data. From the results, we observe that the K-Nearest Neighbors (KNN) method performs reasonably well for classification, but may not be ideal. When we examine the objective values of the estimated labels and compare them with the true labels, we observe that they are quite similar. In other words, although the estimated labels may not match with the true labels, their objective function values are in close proximity. This notion is illustrated in Fig.~\ref{fig:4_nodes_KNN}. The y-axis represents thermal endurance, while the x-axis represents the average heat load for the 25 test cases. For each average heat load value, there are 73 different configurations, denoted by dots in the plot. The configuration estimated as the best configuration is marked with an asterisk ($\ast$). Notably, the asterisk is typically positioned close to the top among all 25 (test population size) cases with different disturbances. Consequently, even though the labeling accuracy is not high (0.7), the objective value of the estimated case aligns closely with that of the optimal case. Thus, the estimated configuration is fairly close to the optimal solution. Furthermore, it is evident that as the mean heat load increases, the thermal endurance decreases, as depicted in the plot.

\begin{table}[ht!]
\small
\centering
\caption{The classification results for different classification techniques with 4 CPHXs}
\label{tab:classiifcation result-4nodes}       
\scalebox{0.72}{
\begin{tabular}{cccc}
\toprule
Method     & Acc (Test/Train)  & Method        & Acc (Test/Train)  \\ \hline
Logistic Regression        & 0.28/0.44 & Random Forest & 0.62/0.69 \\
K-Nearest Neighbours (K-NN)        & 0.70/0.85  & Naïve Bayes   & 0.56/0.62 \\
SVC (Support Vector Classifier)        & 0.24/0.53 & AdaBoost        & 0.24/0.22 \\
Kernel SVM (Support Vector Machine) & 0.56/0.72  & MLP    & 0.73/0.78 \\
Decision Tree   & 0.52/0.57 &            &  \\ \bottomrule
\end{tabular}
}
\end{table}

\section{Results and Discussion}
\label{Results}

The method described in the previous section offers several advantages. Firstly, it enables obtaining the optimal design for heat management systems with varying heat loads without the need to solve the OLOC (Optimal Location of Components) problem. This significantly reduces the computational cost involved. Consequently, engineers can obtain an initial estimate of the optimal design for a given heat load, allowing them to allocate more resources and utilize advanced models to further analyze that specific configuration. Moreover, since the discussed graphs serve as building blocks for larger systems, the acquired knowledge can be leveraged to design configurations with higher complexity. This capability is demonstrated in the subsequent sections. This highlights an additional benefit of conducting optimization studies for new system classes that may lack design heritage—it can expedite the development of engineering knowledge, facilitating the successful design of unprecedented systems.

\subsection{Examination of 8-Node Graphs with Single-Split Branches}
\label{sec: Brute Force Intuition check from extracted knowledge-3nodes}

In Sect.~\ref{sec: CPHx with 3 nodes- Single Split} we endeavored to extract knowledge for small systems having 3 nodes with the goal of applying it to larger systems. These systems can serve as building blocks for larger systems. In this section, we will examine whether this knowledge is generalizable to larger systems. Here, a large graph consisting of 2 junctions, with each junction containing 3 CPHX nodes, is investigated. With 13 different cases possible for the 3 CPHXs in each junction, the large graph encompasses a total of 169 different configurations. This study considers two scenarios with two different disturbances. In the first scenario the disturbance is [5,5,7,5,4,4,4,4] kW, and in the second one the disturbance is [5,5,9,3,2,10,3,2] kW, where the first two elements represent the disturbance of the CPHX mounted on junctions, while the others indicate the disturbance of CPHXs connected to each junction. 

\begin{figure}[ht!]
    \centering
    \subcaptionbox{dist=[5,5,7,5,4,4,4,4] kW}{\includegraphics[width=1.0\linewidth]{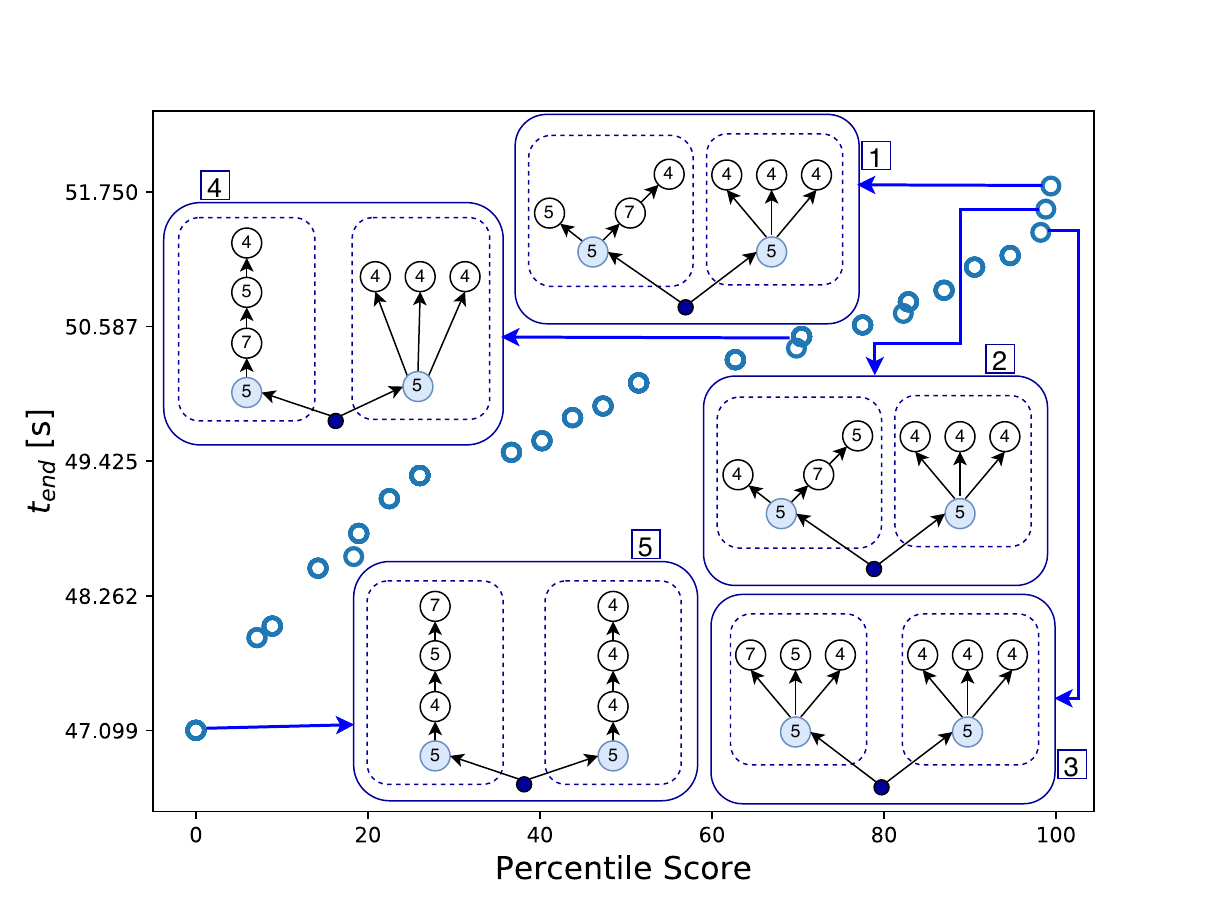}}
    \subcaptionbox{dist=[5,5,9,3,2,10,3,2] kW}{\includegraphics[width=1.0\linewidth]{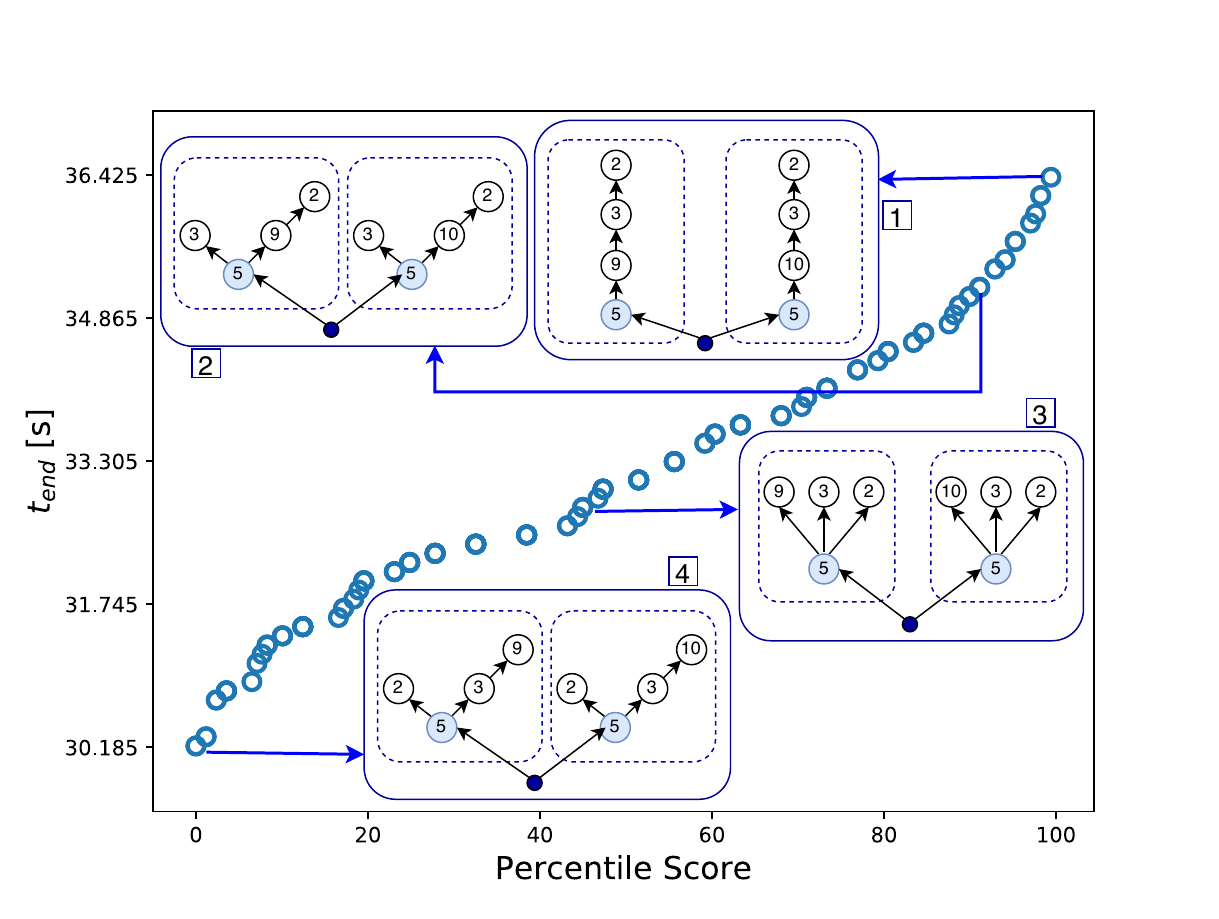}}
    \caption{A case-study to show that the extracted knowledge from Sect.~\ref{sec: CPHx with 3 nodes- Single Split} is applicable to larger systems. The configurations studied in that section are building blocks of the larger systems.}
    \label{fig:Multy_bruteforce_intuition_check}
\end{figure}

Figure~\ref{fig:Multy_bruteforce_intuition_check}(a) presents the results for the first scenario. The y-axis represents thermal endurance, while the x-axis displays the percentile score, which defines the relative position of a specific data point within a dataset by indicating the percentage of data points in the dataset that are equal to or below that particular value. Consequently, in this figure, the top right corresponds to the best solution, while the bottom left represents the worst solution.
In Fig.~\ref{fig:Multy_bruteforce_intuition_check}(a), all three nodes connected to the right junctions have the same heat-load equal to 4 kW. There are 13 possible configurations of these nodes, as shown in Fig.~\ref{fig:Single_3_configs}. In Fig.~\ref{fig:Single_3_configs}, when disturbances are close in value, the best result occurs when all nodes are connected in parallel (configuration 0). This is  what we see for the optimal result in Fig.~\ref{fig:Multy_bruteforce_intuition_check}(a) for case 1. For the left junction, the disturbances are $[7,5,4]$ kW. According to Fig.~\ref{fig:Single_3_configs}, the best configuration for this disturbance is configuration 3 ($(d_1=7)>(d_2=5)>(d_3=4)$), and this is what we see for the the left junction of case 1. 

We see a similar pattern in other cases of  Fig.~\ref{fig:Multy_bruteforce_intuition_check}(a). For example for the worst case (case 5), all of the nodes in the right junction are in series. In this case, there is no control over the flow because the flow rate is equal to the pump flow rate which is a fixed parameter. Hence, there is no optimization problem to solve, and that results in a poor objective value. Additionally, in the left junction, the node with the maximum heat-load is the farthest node from the pump, and the node with lowest heat-load is the closest to the pump. However, based on Fig.~\ref{fig:Single_3_configs}, when all nodes are in series, the best result occurs when the node with maximum load is closest to the pump and the node with the minimum load is the farthest to the pump, and this is the opposite of the case we see in the left junction of case 5, and that is why this has the lowest thermal endurance.

Figure~\ref{fig:Multy_bruteforce_intuition_check}(b) shows the results of the second study where the disturbances are $[5,5,9,3,2,10,3,2]$ kW. The optimal solution for this configuration is case 1. This optimal result is well aligned with the extracted knowledge presented in Fig.~\ref{fig:Single_3_configs}. For the right junction, the disturbances are $[10,3,2]$, and based on Fig.~\ref{fig:Single_3_configs}, the optimal result for this case where ($(d_1=10)>>(d_2=3)>(d_3=2)$) happens for configuration 12, where all nodes are in series and the node with maximum load is closest to the tank, and the node with minimum load is the farthest away from the tank; this is what we see in the right junction of case 1. For the left junction, the disturbances are $[9,3,2]$, and based on Fig.~\ref{fig:Single_3_configs}, the optimal result for this case where ($(d_1=9)>>(d_2=3)>(d_3=2)$) similarly happens for configuration 12. It should be noted that the best configuration is dependent on the disturbances. For example, in Fig.~\ref{fig:Multy_bruteforce_intuition_check}(a), the worst result happens for the case where nodes are in series (case 5 in part a), but in Fig.~\ref{fig:Multy_bruteforce_intuition_check}(b) the best solution is the case where all nodes are in series (case 1 in part b). In addition, the extracted knowledge is also influenced by how the disturbances are related. The results indicate that the knowledge could be used to guide designers to determine which configuration is closer to the optimal solution. Then, the engineer could start designing a thermal management system based on this information. It should be noted that the results in the study are limited and further exploration is required to have a more comprehensive understanding of different scenarios and use cases. 

\subsection{Examination of 9-Node Graphs with Multi-split Branches}
\label{sec: Brute Force Intuition check from extracted knowledge-3_ad_4_nodes}

\begin{figure}[ht!]
    \centering
    \includegraphics[width=1.0\linewidth]{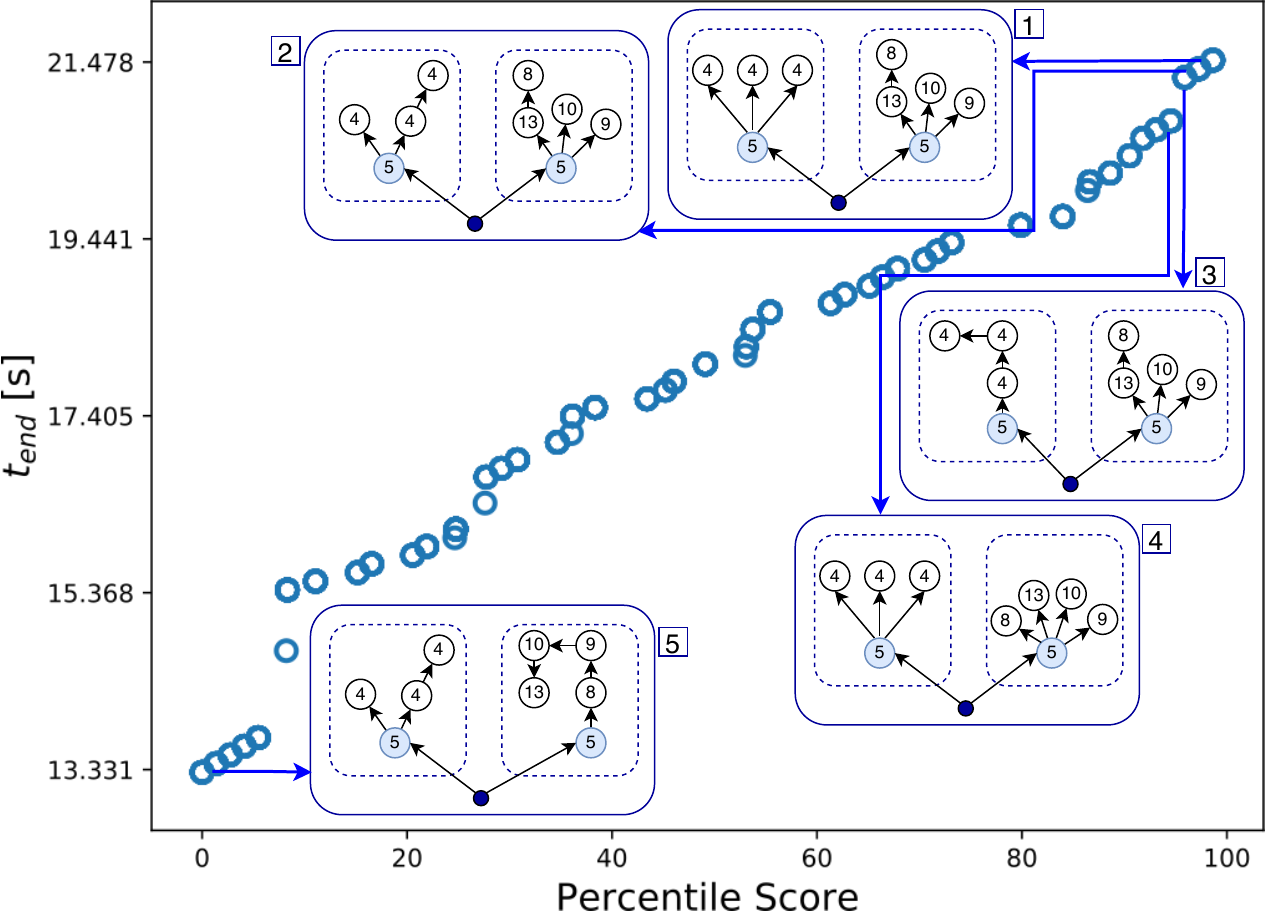}
    \caption{A case-study to show that the combination of extracted knowledge from Sect.~\ref{sec: CPHx with 3 nodes- Single Split} and \ref{sec: CPHx with 4 nodes- Single} is applicable to larger systems. The configurations studied in that section are building blocks of the larger systems.}
    \label{fig:multi_3_and_4}
\end{figure}

In this section, we apply the knowledge obtained from the previous sections to multi-split graphs with 2 junctions. The first junction branches into 3 nodes, while the second junction branches into 4 nodes. The results are depicted in Fig.~\ref{fig:multi_3_and_4}. In the case of the first junction, where all three nodes are connected to it, each node has an identical heat load of 4 kW. Based on the trained classifier, the optimal solution indicates that all nodes should be connected in parallel. Similarly, for the next junction with 4 nodes, the trained classifier suggests that the optimal configuration consists of 3 branches. In one of these branches, the node with the maximum heat load is in series with the node with the minimum heat load, and the node with the maximum heat load is positioned closer to the pump. As shown in the figure, this configuration aligns perfectly with the optimal solution (case 1).

Furthermore, it is worth noting that although case 4 exhibits maximum control authority, with all branches in parallel at each junction, it does not necessarily yield the optimal configuration. Through these studies, we can gain knowledge that may not be achievable through conventional human engineering practices.

\subsection{Automating the Design of Complex Graphs Using the Trained Models}
\label{Automating the Design of Complex Graphs Using Trained Models Derived from Simple Graphs}

In the previous section, we observed that larger graphs can be divided into smaller graphs, and a model trained on these smaller graphs can be used to design larger graphs. This section aims to automate this process. Given a complex, multi-split graph, it first divides the graph into several groups. The groups consist of CPHXs nodes connected to a tank or junctions. Based on the number of CPHXs in each group, the trained model for that group of nodes is used to estimate the optimal configuration for that subgraph. After performing this process for all subgraphs, the optimal design for the complex graph is estimated by combining the subgraphs.

\begin{figure}[ht!]
    \centering
    \includegraphics[width=1.0\linewidth]{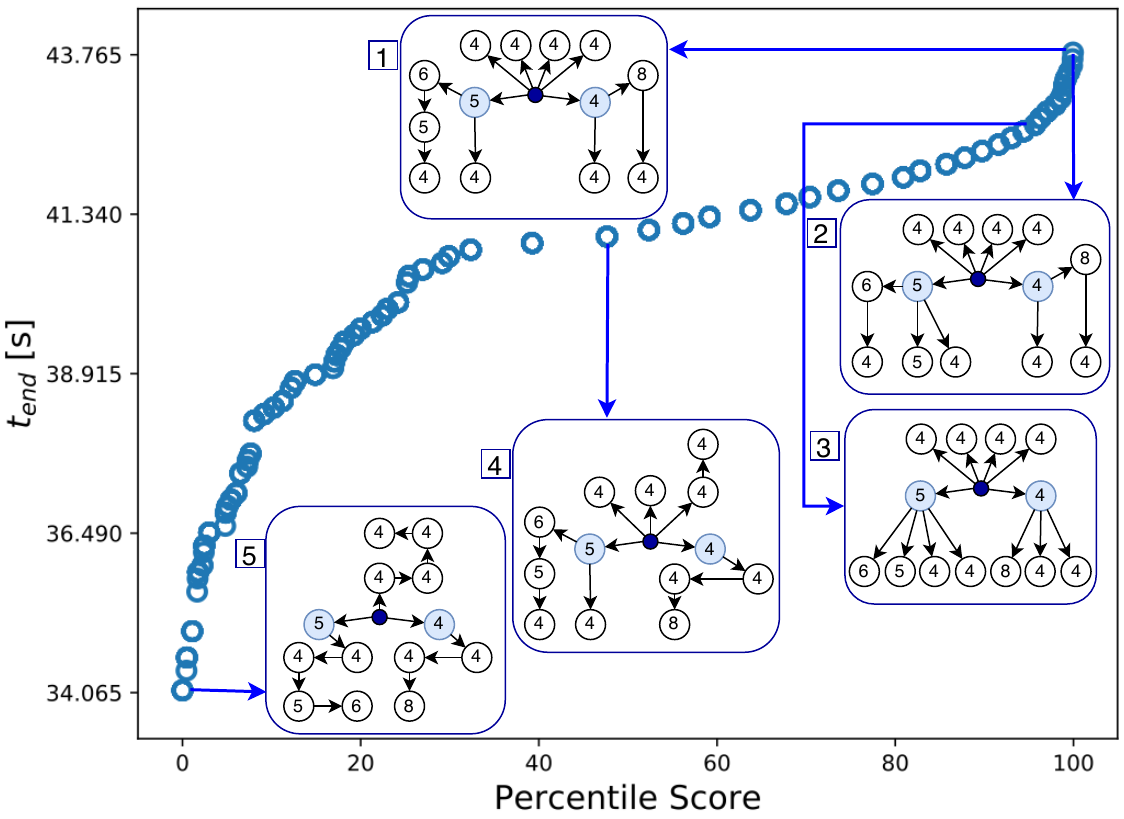}
    \caption{Automated division of complex graphs into subgraphs and estimation of optimal designs using trained models}
    \label{fig:paperB_percentile_cropped}
\end{figure}

The process is illustrated in Figure~\ref{fig:paperB_percentile_cropped}. In this graph, consisting of 13 CPHXs, four nodes are directly connected to the tank, four nodes are connected to a junction, and three nodes are connected to another junction. The heat load of each node is written inside the circles in kilowatts. Enumerating all possible configurations for this complex graph would require examining 69,277 different cases. If instead of three nodes, four nodes were connected to the other junction, this number would increase to 389,017. It is clear that enumeration is not feasible for complex graphs.

However, based on the proposed method, this complex graph is divided into three subgraphs, with two of them having four nodes and the third one having three nodes. Then the trained model is used to estimate the optimal design. Figure~\ref{fig:paperB_percentile_cropped} shows the results of this study, where Case-2 represents the estimated optimal design obtained from this approach. To evaluate the quality of this optimal design, its objective value is compared with 2,000 randomly selected configurations out of the 69,277 cases. As we can observe, Case-2 is quite close to Case-1, which is the best solution among these 2,000 cases. Thus, this approach successfully estimates the optimal complex graph without having to enumerate all 69,277 cases. By using this method, the estimated optimal design can be obtained within a few seconds, whereas enumeration of 69,277 cases would take several days even with parallel computation. In Fig.~\ref{fig:paperB_percentile_cropped}, Cases 3, 4, and 5 show other configurations along with their objective values. The worst-case scenario, Case 5, involves all nodes in series, with the node having the highest heat load farthest from the tank. Interestingly, Case 3, where all nodes within each subgraph are parallel and have maximum control authority, did not yield the best result. This indicates that the graph configuration significantly impacts the dynamics, and having more control authority does not necessarily guarantee better results. This method can be easily applied to graphs with multiple junctions and can estimate the optimal configuration with good accuracy.

\subsection{ Use of Trained KNN with Three CPHXs for Optimal Label Estimation in the Presence of Four CPHXs}
\label{Use of Trained KNN with Three CPHXs for Optimal Label Estimation in the Presence of Four CPHXs}

In the previous sections, we investigated the trained models to determine if their results are generalizable to more complex configurations. Through various examples, we demonstrated that this is indeed the case. Now, in this section, our aim is to investigate whether the trained model with fewer nodes can accurately estimate the optimal class in scenarios with a larger number of nodes. This inquiry is of interest because, if achievable, it would eliminate the need to enumerate all graphs generated with a high number of nodes in order to either select an appropriate configuration or generate training data. For example, the number of different configurations for single split cases with 3, 4, and 5 nodes are 9, 13, and 501, respectively. It is evident that as we increase the number of nodes, the number of graphs increases rapidly, requiring a significant amount of training data to build a model. Alternatively, using brute force would necessitate studying a vast number of cases. Consequently, utilizing pretrained models with fewer nodes to estimate configurations with more nodes would be advantageous.

\begin{figure}[ht!]
    \centering
    \includegraphics[width=0.9\linewidth]{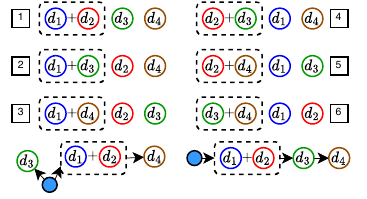}
    \caption{By combining disturbances in graphs with 4 CPHXs, we can generate graphs with 3 CPHXs. This process leads to 6 distinct cases labeled from 1 to 6. In these cases, a dashed box represents a new CPHX in the graph with 3 nodes, and its disturbance is obtained by combining the corresponding disturbances from the graphs with 4 CPHXs. Figure~\ref{fig:Single_3_configs} displays the 13 different graphs that result from using 3 CPHXs. At the bottom of current figure, two sample graphs are provided as illustrations.}
    \label{fig:single_3_for_4_diagram}
\end{figure}

In order to achieve this objective, we utilize the tariend KNN model with 3 CPHXs to determine the optimal class with 4 CPHXs. To accomplish this, we combine the 4 nodes to obtain 3 new nodes. Figure \ref{fig:single_3_for_4_diagram} displays six different load cases that result from combining these 4 nodes. Here, $d_i$ represents the disturbance of the \textit{ith} node. With 4 nodes, we can generate six distinct load cases, each consisting of 3 nodes. For instance, in configuration 1, the disturbances of node 1 and 2 are summed together. Once we have these new configurations, derived from a single graph with 4 nodes, we employ the trained model to determine the best estimated configuration for all 6 load cases shown in Fig.\ref{fig:single_3_for_4_diagram}. 

In this approach, instead of solving 73 configurations for graphs with 4 CPHXs, we only need to consider 6 estimated optimal configurations for graphs with 3 CPHXs. We then compare their objective values and select the best one. In this section, we applied this method to graphs with 4 CPHXs under 200 different disturbances, each consisting of 4 elements (as we have 4 CPHXs). We used the trained model with 3 CPHXs to estimate the optimal class for all 6 cases shown in Fig.~\ref{fig:single_3_for_4_diagram}. Subsequently, we compared their objective values and selected the best configuration among those 6 cases. Then, the equivalent configuration with 4 nodes was obtained, and its objective value was labeled as $\hat{y}$. For each disturbance, out of the 73 different configurations (for graphs with 4 CPHXs), $\hat{y}$ represents the estimated objective value, and the corresponding class indicates the estimated optimal configuration. Among these 73 different configurations, the maximum value is labeled as $y_{\mathrm{max}}$, and the minimum value is labeled as $y_{\mathrm{min}}$. We define a variable as $\frac{y_{\mathrm{max}} - \hat{y}}{y_{\mathrm{max}} - y_{\mathrm{min}}}$. This variable ranges between 1 and 0, where 0 indicates that the estimated objective value is the same as the true optimal value, and 1 indicates that the estimated objective value is the worst among all 73 populations

\begin{figure}[ht!]
    \centering
    \subcaptionbox{}{\includegraphics[width=1.0\linewidth]{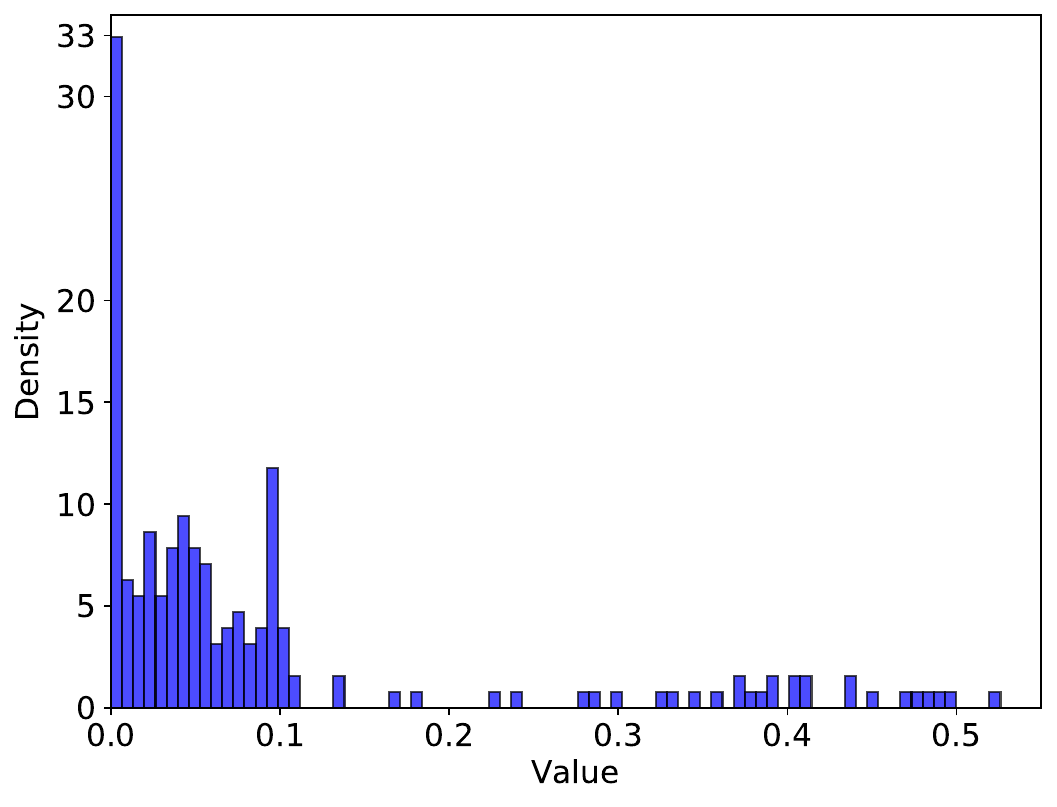}}
        \subcaptionbox{}{\includegraphics[width=1.0\linewidth]{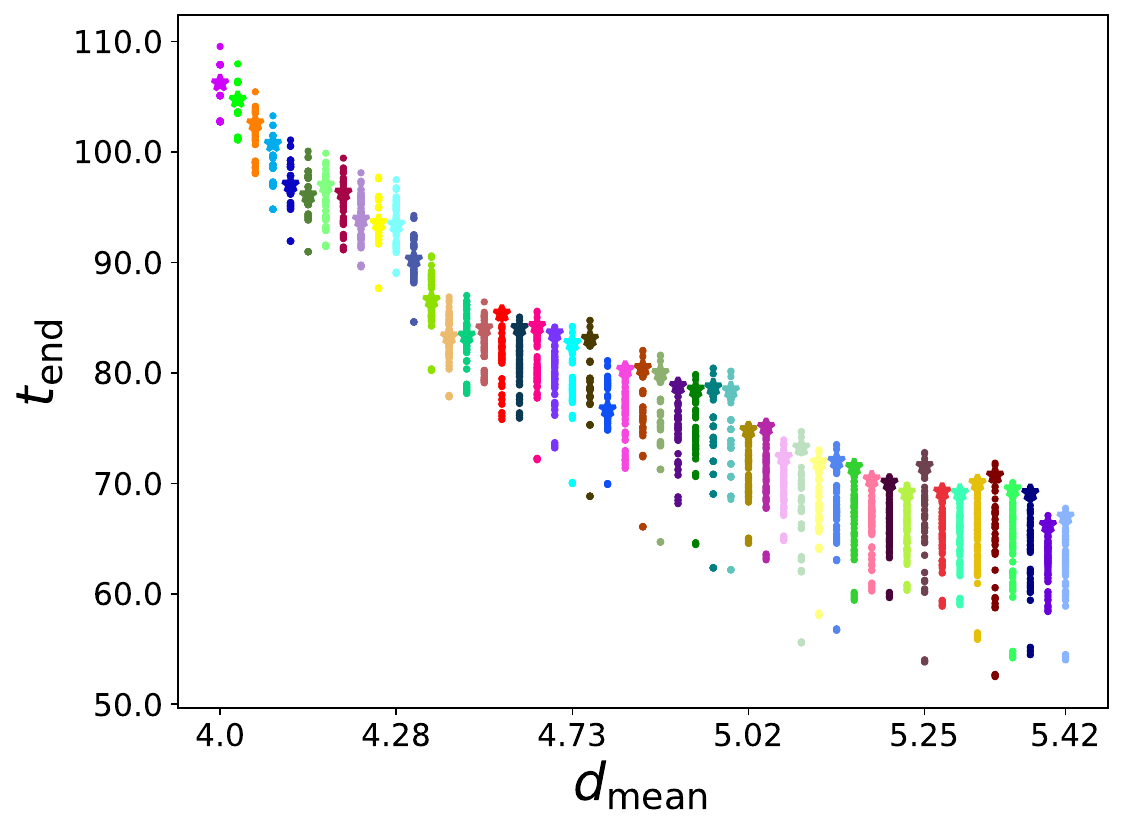}}
    \caption{(a): The histogram represents a population of 200 different disturbances, each containing 4 elements (as we have 4 CPHXs). The x-axis represents $\frac{y_{\mathrm{max}} - \hat{y}}{y_{\mathrm{max}} - y_{\mathrm{min}}}$, and the y-axis represents the density. A lower value on the x-axis indicates a better result, as it signifies that the objective value of the estimated configuration is closer to that of the optimal configuration. (b): Objective value comparison of estimated optimal configuration ($\ast$) with Other populations ($\cdot$) for each disturbance.}
    \label{fig:single_3_for_4}
\end{figure}

The histogram in Fig.~\ref{fig:single_3_for_4}(a) illustrates the results. The x-axis represents $\frac{y_{\mathrm{max}} - \hat{y}}{y_{\mathrm{max}} - y_{\mathrm{min}}}$, while the y-axis denotes the density. From the plot, we observe that among the 200 different disturbances, the estimated class (and its corresponding objective value) tends to be close to the optimal solution, as indicated by the x-axis shifting towards 0. However, it should be noted that we should not expect high accuracy in this estimation process, as we utilized trained models with fewer nodes to estimate the optimal class for configurations with more nodes. In Fig.~\ref{fig:single_3_for_4}(b), the estimated objective value ($\ast$) is also shown along with the other 73 configurations ($\cdot$). We observe that in low disturbance, the estimated optimal objective value falls in the middle of the 73 configurations, indicating relatively lower accuracy for this method. However, as we move towards higher disturbance (to the right), the accuracy generally improves.

It should be noted that if two CPHXs are arranged in series, the result is not simply equivalent to combining two nodes and summing their disturbances. This is because the combined node with two disturbances has more interaction with the environment (more surface area exposed to the environment) compared to a single node with its disturbance being the sum of the two individual nodes. To obtain an equivalent configuration, adjustments need to be made to the heat exchanger characteristics of the combined node.

\section{Limitations}
\label{Limitations}

While the method discussed in this article offers certain benefits, it also has limitations that should be considered. Firstly, when dealing with graphs containing a larger number of nodes, additional training data is required. As the number of nodes increases, the number of configurations grows exponentially. For instance, increasing the number of nodes from 4 to 5 results in a jump from 72 to 501 configurations. 
To generate the training data, the corresponding OLOC problems must be solved across a range of different heat loads. For a specific heat load and configuration, solving the corresponding OLOC problem takes approximately 2 minutes using the specified workstation with AMD EPYC 7502 32-Core Processor @ 2.5 GHz, 64 GB DDR4-3200 RAM, LINUX Ubuntu 20.04.1, and Python 3.8.10. Even with parallel computation using 64 cores, solving the OLOC problem for 501 different configurations would still take about 16 ($ \approx 501/64*2$) minutes. If the heat load population size is 200, the computation time would extend to approximately 53 hours ($ \approx 200*16/60$). Although this might still be reasonable, as the number of graphs increases, the computation time will grow exponentially.
Consequently, applying this method to graphs with a large number of nodes may not be feasible. To address this issue, it has been demonstrated that dividing the graphs into sections based on junctions and the tank, utilizing the trained model for each section, and then combining the results can help estimate the optimal configuration of the original complex system. However, this approach relies on the assumption that the subgraphs are relatively simple, typically comprising 3 or 4 nodes. An alternative approach to tackle this challenge could be to employ optimization algorithms like Genetic Algorithm (GA) to estimate the best configuration, rather than generating extensive training data for large graphs.

Secondly, the accuracy of estimating the optimal configuration using the discussed method is not particularly high, approximately 75\%. To improve this accuracy, alternative machine learning algorithms with different features can be employed. Nonetheless, the paper demonstrates that despite the modest accuracy in estimating the optimal configuration, the difference between the objective value of the estimated configuration and the best configuration is relatively small. As a result, this method can be employed as an initial estimate for determining the optimal configuration.

The future work will address these limitations by utilizing population-based algorithms, such as GA, to identify the connections between tanks, junctions, and CPHXs that lead to better objective values. Additionally, a more comprehensive study will be conducted on feature selection and hyperparameter tuning of the machine learning methods employed in this research.

\section{Conclusion}
\label{Conclusion}
This article introduces the knowledge extraction process for fluid-based thermal management systems with single-split and multi-split configurations. The configurations are generated using a graph-based modeling technique, considering a range of different disturbances. The open-loop optimal control approach is employed to address the optimization problem for each configuration. Subsequently, various machine learning techniques are applied to extract knowledge from the data, and the trained model's accuracy is evaluated for the objective function value (thermal endurance) and optimal class label prediction using both training and test data.

The trained model is then tested for its applicability in designing new and complex heat management systems. The results demonstrate that the trained model can be effectively generalized to design new, intricate single-split and multi-split fluid-based thermal management systems. One significant advantage of the trained model is that it provides designers with an initial estimate of the optimal class without having to solve the open-loop optimal control problem. Armed with this initial estimate, designers can employ more advanced models to further evaluate the specific configuration. This enhances the engineering design process for fluid-based thermal management systems.

Furthermore, this approach offers two notable advantages by not relying solely on previous human-centered data. Firstly, it can be used to design new systems that lack significant design heritage. Secondly, the training data used in this method are optimized, whereas previous human designs may not necessarily be optimal. Consequently, this approach contributes to improving the engineering design of such systems. Moreover, this methodology has the potential for application in diverse engineering problems, offering similar opportunities in other disciplines. The limitations of the present work has been discussed and suggestions for future work are also made.

\bibliographystyle{asmejour}   

\bibliography{Main} 

\begin{thebibliography}{10}
\newcommand{\enquote}[1]{``#1''}
\providecommand{\url}[1]{\texttt{#1}}
\providecommand{\urlprefix}{}
\expandafter\ifx\csname urlstyle\endcsname\relax
  \providecommand{\doi}[1]{doi:\discretionary{}{}{}#1}\else
  \providecommand{\doi}{doi:\discretionary{}{}{}\begingroup \urlstyle{rm}\Url}\fi
\providecommand{\eprint}[2][]{\urlprefix\url{#1#2}}
\providecommand{\hrefurl}[2][]{\href{#1}{#2}}

\bibitem{buidin2021battery}
Buidin, T. I.~C. and Mariasiu, F., 2021, \enquote{Battery thermal management systems: Current status and design approach of cooling technologies,} Energies, \textbf{14}(16), p. 4879.

\bibitem{bayat2023multisplit}
Bayat, S., Shahmansouri, N., Peddada, S.~R., Tessier, A., Butscher, A., and Allison, J.~T., 2023, \enquote{Multi-split configuration design for fluid-based thermal management systems,} \eprint{2310.15500}

\bibitem{mathew2022review}
Mathew, J. and Krishnan, S., 2022, \enquote{A review on transient thermal management of electronic devices,} Journal of Electronic Packaging, \textbf{144}(1).

\bibitem{liu2021review}
Liu, H., Wen, M., Yang, H., Yue, Z., and Yao, M., 2021, \enquote{A review of thermal management system and control strategy for automotive engines,} Journal of Energy Engineering, \textbf{147}(2), p. 03121001.

\bibitem{laloya2015heat}
Laloya, E., Lucia, O., Sarnago, H., and Burdio, J.~M., 2015, \enquote{Heat management in power converters: From state of the art to future ultrahigh efficiency systems,} IEEE transactions on Power Electronics, \textbf{31}(11), pp. 7896--7908.

\bibitem{shah2005exergy}
Shah, A.~J., Carey, V.~P., Bash, C.~E., and Patel, C.~D., 2005, \enquote{Exergy-based optimization strategies for multi-component data center thermal management: Part I—analysis,} \textit{International Electronic Packaging Technical Conference and Exhibition}, Vol. 42002, pp. 205--213.

\bibitem{aloui2003handbook}
Aloui, F. and Groll, E.~A., 2003, \textit{Handbook of Thermal Management Systems}, Elsevier.

\bibitem{jafari2018thermal}
Jafari, S. and Nikolaidis, T., 2018, \enquote{Thermal management systems for civil aircraft engines: Review, challenges and exploring the future,} Applied Sciences, \textbf{8}(11), p. 2044.

\bibitem{SaeidBayat-Vehicle}
Bayat, S. and Allison, J.~T., 2023, \enquote{{Control Co-Design} With Varying Available Information Applied to Vehicle Suspensions,} \textit{International Design Engineering Technical Conferences and Computers and Information in Engineering Conference}, American Society of Mechanical Engineers.

\bibitem{bayat2023nested}
Bayat, S., Lee, Y.~H., and Allison, J.~T., 2023, \enquote{Nested Control Co-design of a Spar Buoy Horizontal-axis Floating Offshore Wind Turbine,} \eprint{2310.15463}

\bibitem{Belm2018FromDT}
Bel{\'e}m, C., 2018, \enquote{From Design to Optimized Design - An algorithmic-based approach,} Proceedings of the 36th International Conference on Education and Research in Computer Aided Architectural Design in Europe (eCAADe) [Volume 2].

\bibitem{fuge2014machine}
Fuge, M., Peters, B., and Agogino, A., 2014, \enquote{{Machine Learning Algorithms for Recommending Design Methods},} Journal of Mechanical Design, \textbf{136}(10), p. 101103.

\bibitem{holzinger2019introduction}
Holzinger, A., 2019, \enquote{Introduction to MAchine Learning \& Knowledge Extraction (MAKE).} Mach. Learn. Knowl. Extr., \textbf{1}(1), pp. 1--20.

\bibitem{MacDonald2020ExplainingNN}
MacDonald, J., W{\"a}ldchen, S., Hauch, S., and Kutyniok, G., 2020, \enquote{Explaining Neural Network Decisions Is Hard,} .

\bibitem{Liu2018ImprovingTI}
Liu, X., Wang, X., and Matwin, S., 2018, \enquote{Improving the Interpretability of Deep Neural Networks with Knowledge Distillation,} 2018 IEEE International Conference on Data Mining Workshops (ICDMW), pp. 905--912.

\bibitem{peddada2019optimal}
Peddada, S.~R., Herber, D.~R., Pangborn, H.~C., Alleyne, A.~G., and Allison, J.~T., 2019, \enquote{Optimal flow control and single split architecture exploration for fluid-based thermal management,} Journal of Mechanical Design, \textbf{141}(8).

\bibitem{falck2021dymos}
Falck, R., Gray, J.~S., Ponnapalli, K., and Wright, T., 2021, \enquote{dymos: A Python package for optimal control of multidisciplinary systems,} Journal of Open Source Software, \textbf{6}(59), p. 2809.

\bibitem{gill2005snopt}
Gill, P.~E., Murray, W., and Saunders, M.~A., 2005, \enquote{SNOPT: An SQP algorithm for large-scale constrained optimization,} SIAM review, \textbf{47}(1), pp. 99--131.

\bibitem{biegler2009large}
Biegler, L.~T. and Zavala, V.~M., 2009, \enquote{Large-scale nonlinear programming using IPOPT: An integrating framework for enterprise-wide dynamic optimization,} Computers \& Chemical Engineering, \textbf{33}(3), pp. 575--582.

\bibitem{bayat2023ss}
Bayat, S. and Allison, J.~T., 2023, \enquote{SS-MPC: A user-friendly software based on single shooting optimization to solve Model Predictive Control problems,} Software Impacts, \textbf{17}, p. 100566.

\bibitem{bayat2023lgrmpc}
Bayat, S. and Allison, J.~T., 2023, \enquote{LGR-MPC: A user-friendly software based on Legendre-Gauss-Radau pseudo spectral method for solving Model Predictive Control problems,} \eprint{2310.15960}

\end{thebibliography}



\end{document}